\documentclass[epj]{webofc}
\usepackage[utf8]{inputenc}
\usepackage[varg]{txfonts}   
\usepackage{booktabs}
\usepackage{xcolor}
\definecolor{darkred}{rgb}{0.4,0.0,0.0}
\definecolor{darkgreen}{rgb}{0.0,0.4,0.0}
\definecolor{darkblue}{rgb}{0.0,0.0,0.4}
\usepackage[bookmarks,linktocpage,colorlinks,
    linkcolor = darkred,
    urlcolor  = darkblue,
    citecolor = darkgreen]{hyperref}
\def\e{{\rm e}}
\def\tr{{\rm tr}}

\newcommand{\C}{\mathbb{C}}

\newcommand{\bR}{\mathbb{R}}

\renewcommand{\Re}{\rm Re}
\renewcommand{\Im}{\rm Im}

\newcommand{\cG}{{\cal G}}

\newcommand{\cM}{{\cal M}}
\newcommand{\cO}{{\cal O}}

\def\DD{{D\hskip-6pt \slash}}
\def\be#1\ee{\begin{equation}#1\end{equation}}
\def\bea#1\eea{\begin{align}#1\end{align}}
\def\bra{\langle}
\def\ket{\rangle}
\wocname{EPJ Web of Conferences}
\woctitle{Lattice2017}
\begin{document}
%
\selectlanguage{english}
\title{%
Status of Complex Langevin
}
\author{%
\firstname{Erhard} \lastname{Seiler}\inst{1}\fnsep\thanks{Plenary Lecture 
given at Lattice 2017, the 35th International Symposium on Lattice Field 
Theory, Granada, Spain, 18–24 June 2017. The author gratefully 
acknowledges many years of collaboration on these matters with G.~Aarts, 
D.~Sexty and I.-O.~Stamatescu, as well as financial support by the 
organizers of Lattice 2017 and Deutsche Forschungsgemeinschaft (DFG).
I am also grateful to L.~L.~Salcedo for useful remarks.}
\fnsep\thanks{\e-mail: {ehs@mpp.mpg.de}}
}
\institute{%
Max-Planck-Institut f\"ur Physik (Werner-Heisenberg-Institut)\\
F\"ohringer Ring 6\\
80805 M\"unchen\\
Germany
}
\abstract{%
I review the status of the Complex Langevin method, which was invented to
make simulations of models with complex action feasible. I discuss the
mathematical justification of the procedure, as well as its limitations
and open questions. Various pragmatic measures for dealing with the
existing problems are described. Finally I report on the progress in the
application of the method to QCD, with the goal of determining the phase
diagram of QCD as a function of temperature and baryonic chemical
potential.
}
\maketitle
\section{Introduction}\label{sec-1}

\subsection{Sign problem}\label{sec-2}

It is well known that in many instances the functional measure in 
Euclidean Quantum Field Theory is not positive, making standard 
importance sampling impossible. Examples are
\begin{itemize}
\item
The real time Feynman path integral,
\item
theories with topological terms or a nonzero vacuum angle $\theta$,
\item
theories with nonzero chemical potential corresponding to nonzero density.
\end{itemize}

In such cases the functional measure is described by a complex (or 
nonpositive) density $\rho$ on a real configuration space $\cM$. The 
following motto describes a strategy to deal with this `sign problem':
\vskip2mm
\begin{center}
{\em \large When the measure is complex, complexify the fields}
\end{center} 
\vskip2mm
This means the following: for holomorphic observables $\cO$, one tries to 
represent the complex measure by a measure on the complexified 
configuration space $\cM_c$ which is either nonnegative 
or at least only mildly oscillating, making the sign problem less severe. 
If that measure is described by a density $P$, consistency requires that
for all holomorphic observables $\cO$
\be
\bra \cO\ket \equiv \int_{\cM} \cO \rho d\mu= \int_{\cM_c} \cO P 
d\mu_c \,,
\label{match}
\ee
where $d\mu, d\mu_c$ are suitable nonnegative a priori measures.

\subsection{Possible solutions}\label{sec-3}

It is obvious that the conditions Eq.(\ref{match}) leave $P$ highly 
underdetermined, since they only restrict the expectation values of 
holomorphic observables, so there are many possibilities for solving it. 
Among those the following have been studied:

\begin{itemize}

\item 
Solve underdetermined problem directly, as proposed by L.~L.~Salcedo in 
various publications since 1993 
\cite{Salcedo:1996sa,Salcedo:2007ji,Salcedo:2015jxd,Salcedo:thiscontrib287}; 
Weingarten \cite{Weingarten:2002xs} gave some general conditions for the 
existence of a solution. A general construction for abelian systems is 
presented in \cite{Seiler:2017vwj}, see also \cite{Seiler:thiscontrib34, 
Wrzykowski:thiscontrib394, Ruba:thiscontrib352}.

\item
The saddle point method generalized to `Lefschetz thimbles' and related 
modifications of the integration path; see for instance 
\cite{Cristoforetti:2012su,Alexandru:2015sua,Mori:2017pne,  
Tanizaki:thiscontrib49,di Renzo:thiscontrib131,
Nishimura:thiscontrib149,Tsutsui:thiscontrib173, Ohnishi:thiscontrib336,
Bedaque:thiscontrib419}; here a residual, probably milder sign problem 
remains.

\item
The complex Langevin (CL) method invented by Parisi \cite{Parisi:1984cs} 
and Klauder \cite{Klauder:1983nn} long ago.
\end{itemize}

Here I will concentrate on the last option. Its advantages are its 
flexibility and straightforward applicability. It has, however also some 
problems, which will be discussed and which have not yet completely 
solved. 

\section{Complex Langevin}\label{sec-4}

After the method was proposed in 1983 by G.~Parisi \cite{Parisi:1984cs} 
and independently by J.~Klauder \cite{Klauder:1983nn}, the 1980s and 1990s 
saw many studies, sometimes showing success, but sometimes the method 
failed to reach convergence and sometimes, even worse, it converged, but 
to an incorrect limit \cite{Ambjorn:1985cv,Ambjorn:1986fz} not satisfying
Eq.(\ref{match}).

This negative finding almost killed interest in the method, but the 
interest was rekindled by a paper J.~Berges and I.-O.~Stamatescu 
\cite{Berges:2005yt}, which showed some success in computing correlation 
functions of a quantum field theory at {\em real time}.

\subsection{Recall real Langevin}\label{sec-5}

Real Langevin, also known as {\em Stochastic Quantization} was proposed 
by G.~Parisi and Y.-S.~Wu in 1981 \cite{Parisi:1980ys} and developed 
further in particular by Batrouni et al \cite{Batrouni:1985jn} ; a 
comprehensive review is due to Damgaard and H.~H{\"u}ffel 
\cite{Damgaard:1987rr}. It should not be confused with {\em Stochastic 
Quantum Mechanics}, invented already in 1966 by E.~Nelson 
\cite{Nelson:1966sp} (for a brief pedagogic review and comparison of both 
approaches see \cite{Seiler:1984pp}).
 
The principle is the following: consider a real action $S(\vec x)$ on some 
configuration space $\cM$ (for simplicity we assume $\cM=\bR^N$). Then the 
real Langevin equation is
\be
d\vec x=\vec K dt+d\vec w, \quad \vec K=-\vec\nabla S\,,
\ee
where $d\vec w$ is the increment of $N$ dimensional Brownian motion;
the corresponding Fokker-Planck equation, describing the time evolution of 
the probability density $P$ is
\be
\dot P(\vec x;t)=L^T P(\vec x;t)\,;\quad L^T\equiv\vec\nabla 
\left(\vec\nabla+(\vec\nabla S)\right)\,.
\ee
Under rather general conditions it can be shown that $P$ converges to the
unique invariant probability density 
\be
P_\infty\propto\exp(-S)\,;
\ee
the crucial conditions for this to hold are
\begin{itemize}
\item
$\rho\equiv\exp(-S)$ is integrable,
\item 
the process is {\em ergodic}.
\end{itemize}
(Ergodicity means that there is only one measure invariant under the time 
evolution.)

The justification of the real Langevin method is well known:
By a similarity transformation $-L^T$ is converted into a positive 
semidefinite operator
\bea
& H_{FP}\equiv \exp(-S/2)(- L^T) \exp(S/2)=\notag\\
&-\left(\vec\nabla+\frac{1}{2} (\vec\nabla S)\right)
\left(\vec\nabla+\frac{1}{2} (\vec\nabla S)\right)\ge 0\,.
\eea
So the spectrum of $L^T$ lies on the positive real axis. $H_{FP}$ has a 
unique ground state $|0\ket$ if and only if the process is ergodic. Then
\bea   
\lim_{t\to\infty} \exp(-H_{FP}t)&= |0\ket \bra 0 |;\notag\\
\lim_{t\to\infty}P(\vec x;t)&\propto \e^{-S(\vec x)}\,.
\eea
But there are {\em counterexamples}, as we will see below! They are 
related to zeroes of $\rho=\exp(-S)$, which can of course mean that 
$S$ has branch points and the drift $\vec K$ has poles.

\subsection{Go complex!}\label{sec-6}

If $S$ is complex, Klauder and Parisi simply postulate stochastic 
equations of the same form as before:
\be
d\vec z=\vec K dt+d\vec w, \quad \vec K=-\nabla \vec S\,,
\ee
where $d\vec w$ is still the real Wiener increment, i.e.
$\vec dw=\vec \eta(t)dt$ where $\vec\eta$ is white noise with covariance  
$\bra \eta(t)\eta(t')\ket=2 \delta(t-t')$. Obviously the trajectories of 
the process will wander into the complexification $\cM_c=\C^N$. The 
process, however, is still a real stochastic process, but now  on 
$\cM_c$, as seen by writing it for the real and imaginary parts:
\bea
d\vec x=&\vec K_x dt+ d\vec w,\quad \vec K_x=\Re\,\vec K, \notag\\
d\vec y=&\vec K_y dt,\quad\quad\quad\;\, \vec K_y=\Im\,\vec K\,. 
\eea
But the unavoidable question is: Why should this be right?  

\subsection{Justification of Complex Langevin}\label{sec-7}

Early attempts \cite{Nakazato:1986dq,Okano:1992hp} tried to justify 
the method using a relation involving analytic continuation of $P$ in the 
form $\exp(i\vec y \cdot \partial_{\vec x}) P(\vec x,\vec y;t)$: requiring 
consistency by
\be
\int d\vec y \exp(i\vec y \cdot \partial_{\vec x}) P(\vec x,\vec y;t)
=\rho(\vec x;t)
\ee
and taking time derivatives. One problem here is that in general one 
does not know if $P$ has the analyticity of $P$ needed to make sense of 
the left hand side of this equation, espectially since we typically want 
to start with $P(\vec x, \vec y;0)=\delta(\vec x)\delta(\vec y)$.

Our approach \cite{Aarts:2009uq} to the justification is different: let 
$\cO$ be a holomorphic `observable' of which we want to find the average 
with the density $\rho$, and assume that the drift $\vec K=(\vec \nabla 
\rho)/\rho$ is at least {\em meromorphic}.
 
The evolution of the observable averaged over the process 
\be
\cO(\vec z;t)\equiv \bra\cO(\vec z(t))\ket
\ee
is then (according to Ito calculus) governed by the {\em Langevin 
operator} $L$:
\be
\partial_t \cO(\vec z;t)= L\cO(\vec z;t)\,,\quad
L\equiv\left[\vec\nabla_x+\vec K_x\right]\cdot \vec \nabla_x+
\vec K_y\cdot\vec\nabla_y\,,\quad \vec z=\vec x+i\vec y\,,
\ee
which has the formal solution
\be
\cO(\vec z;t)\equiv\exp\left[t L\right] \cO(\vec z;0)\,.
\ee
$\cO(\vec z;t)$ will be holomorphic wherever $\vec K$ is, i.e. away from 
the poles of $\vec K$, so there the Cauchy-Riemann (CR) equations hold:
\be
\vec\nabla_y\cO(\vec z;t)=i\vec \nabla_x\cO(\vec z;t)\,.
\ee

The positive density $P$ on $\cM_c$ evolves according to a Fokker-Planck 
equation
\be
\frac{\partial}{\partial t} P(x,y;t)= L^T P(x,y;t)\,;\quad
P(x,y;0)=\delta(x-x_0)\delta(y)\,,
\label{realFPE}
\ee
$L^T\equiv\nabla_x[\nabla_x-K_x]-\nabla_y  K_y$  is the {\em real}
Fokker-Planck operator.

The complex density $\rho$, on the other hand, evolves according to
\be
\label{complexFPE}
\frac{\partial}{\partial t} \rho(x;t)= L_c^T \rho(x;t)\,;\quad
\rho(x;0)=\delta(x-x_0)\,,
\ee
where $L^T_c\equiv \vec\nabla_x \left[\vec \nabla_x - K\right]$ is the 
{\em complex} Fokker-Planck operator, which can be naturally extended to 
act on functions on $\cM_c$. 

The question is then whether the two evolutions are consistent in the 
sense that they lead to identical evolutions of expectation values for 
holomorphic observables, i.e. whether
\be
\bra\cO \ket_{P(t)}\equiv \int \cO(x) P(x,y;t) dx\, dy\,,\qquad
\bra \cO\ket_{\rho(t)}\equiv \int \cO(x) \rho(x;t) dx
\label{twoevol}
\ee
remain equal if they agree at $t=0$. Consider the two differential 
equations determining the two evolutions Eq.(\ref{twoevol}):
\bea
\partial_t \bra \cO\ket_{\rho(t)}&=\int dx \cO(x)\, L_c^T\rho(x;t)
\notag\\
\partial_t \bra \cO\ket_{P(t)}&= \int dx dy\cO(x+iy) L^T P(x,y;t)\,;
\eea 
it can be shown that the right hand sides agree if we can use integration 
by parts without any boundary terms \cite{Aarts:2009uq}. So that would 
ensure the desired consistency. A crucial role in the argument is played 
by the CR equation obeyed by $\cO(\vec z;t)$.

Let us list the obvious assumptions that were used:
\begin{itemize}
\item
agreement of initial conditions  
(irrelevant for $t\to\infty\,$ if the process is ergodic),
\item
meromorphy of drift $\vec K\equiv \vec K_x+i\vec K_y$,
\item
sufficient decay of $|\vec K P\cO|$ at imaginary infinity
and near poles of $\vec K$: this has to be checked.
\end{itemize}
Under these assumption we thus have
\be\boxed{\bra \cO\ket_{\rho(t)}= \bra \cO\ket_{P(t)} \quad
\forall\,t\ge 0}\,.
\ee

We sketch the idea of the proof:

\noindent
1. The initial conditions agree

\noindent
2. Let $\cO(x+iy;t)\equiv\exp\left[t L\right] \cO(x+iy)$ be the (unique) 
solution of the PDE
\be\
\partial_t \cO(x+iy;t)= L\cO(x+iy;t) \quad (t\ge 0)\,;
\ee

\noindent
3. $F(t,\tau)\equiv \int P(x,y;t-\tau) \cO(x+iy;\tau)$:
interpolates between $\bra \cO\ket_{P(t)}$ and
$\bra \cO \ket_{\rho(t)}$:
\be   
F(t,0)= \bra \cO\ket_{P(t)};\quad F(t,t)= \bra \cO \ket_{\rho(t)}
\ee
(the last eqation involves integration by parts).

Formally the quantity $F(t,\tau)$ is independent of $\tau$:
\bea
\frac{\partial}{\partial \tau} F(t,\tau) =
-&\int L^T P(x,y;t-\tau)\cO(x+iy;\tau)dx dy\notag\\
+&\int  P(x,y;t-\tau) L \cO(x+iy;\tau)dx dy\,;
\eea
Integration by parts and the holomorphy of $\cO(\vec z;t)$ imply 
\be
\frac{\partial}{\partial \tau} F(t,\tau)=0\;\quad
\Longrightarrow\quad
\bra \cO\ket_{\rho(t)}= \bra \cO\ket_{P(t)}\,.
\ee
We assumed again that in the integration by parts there are no boundary 
terms at $\infty$ and at the poles of $\vec K$ .

In equilibrium we obtain the so-called {\em consistency conditions} (CC)
\be
\partial_t \bra \cO\ket=\bra L\cO\ket\equiv\int P(x,y;\infty)
L\cO(x+iy)dx\,dy=0\,,
\quad({\rm CC})
\label{cc}
\ee
expressing the stationarity of observables averaged over the 
process. These relations are equivalent to the Schwinger-Dyson equations 
\cite{Aarts:2011ax}. If integration by parts without boundary terms is 
allowed, they are also equivalent to
\be
L^TP=0\,,
\ee
i.e. they express the fact that the equilibrium measure solves the 
stationary real Fokker-Planck equation.

But together with some additional conditions the CC Eq.(\ref{cc}) are 
indeed {\em sufficient} to ensure correctness of the equilibrium measure 
(see \cite{Aarts:2011ax}):

{\bf Theorem:} Consider a compact configuration space $\cM$. Assume that
\begin{itemize}
\item 
Eq.(\ref{cc}) holds for a dense (in supremum norm on $\cM$) set of 
observables $\cO$,
\item
$\left|\int_{\cM_c} P\cO\right|\le const \sup_{\cM} \left|\cO\right|$ for 
all $\cO$ in that dense set,
\item
$0$ is a nondegenerate eigenvalue of $L_c^T$ .
\end{itemize}
Then the equilibrium measure of the CL process is {\em correct}, i.e.
\be
\int_{\cM_c} P\cO= \frac{1}{Z} \int_{\cM} \e^{-S}\cO\,.
\ee 
The hardest point to check is the bound (second bullet point), but in some 
cases one can obtain negative evidence by its violation 
\cite{Aarts:2011ax,Aarts:2017vrv}.

\section{Problems swept under the rug}\label{sec-8}

\subsection{Mathematical problems}\label{sec-9}

There are two serious mathematical problems connected with the CL method:

(1) Existence and uniqueness of the stochastic process as well as 
the time evolutions generated by $L,L_c,L^T,L_c^T$ for all $t\ge 0$ are 
not known. In numerical applications, however, there never seems to be a 
problem, provided one uses a suitable {\em adaptive step size} as 
described in \cite{Aarts:2009dg}. 

(2) Convergence of the positive density $P$ to an equilibrium measure is 
not proven mathematically. What we would need is some information about 
the spectrum of $L$ and $L^T$, namely that the spectrum is contained in 
the left half of the complex plane, with a unique eigenvalue at the 
origin.

Already in 1985 Klauder and Peterson \cite{Klauder:1985ks} remarked about 
the {\em ``conspicuous absence of general theorems''} for such 
non-selfadjoint and non-normal operators. To my knowledge, the statement 
is still valid today.

We should, however, be pragmatic and not be deterred by this lack of 
mathematical rigor: numerically it seems that an equilibrium 
measure exists in all interesting cases; a necessary condition is the 
existence of an attractive fixed point of the drift $\vec K$, which holds 
in all cases of physical interest.

\subsection{Practical problems}\label{sec-10}

Being pragmatic, we still have to worry about some problems. These are:

(1) Boundary terms at $\infty$ and at the poles of $\vec K$ (if present),

(2) Lack of ergodicity.

Both problems may lead to failure of the CL method, as we will show below. 
The simulations have to be monitored for possible boundary terms.

Other authors have also formulated criteria for failure, for instance 
\cite{Hayata:2015lzj}. These criteria are interesting, but it seems they 
can also be subsumed under those mentioned above. Salcedo 
\cite{Salcedo:2016kyy} proves constraints on the support of the positive 
measure on $\cM_c$ that allow in some simple models to predict failure of 
CL without the need to actually carry out a simulation.

\section{Boundary terms at $\infty$}\label{sec-11}

In typical cases of lattice gauge theories the configuration space $\cM$ 
is compact whereas its complexification $\cM_c$ is not, for example
\be
\cM=SU(N)^{\times k},\, \cM_c=SL(N,\C)^{\times k}\,,
\ee
\noindent
where $k$ is the number of links. It is a well-known fact that holomorphic 
functions grow at $\infty$, hence the drift $\vec K$ as well as the 
observables will grow at $\infty$. This can lead to ``skirts'' or 
``tails'' of the distribution of $P$, $|\vec K \cO P|$ on $\cM_c$.

The presence or absence of boundary terms at $\infty$ under integration by 
parts will depend on the decay of the distribution of $|\vec K\cO P|$ 
as we move towards $\infty$. 

In some toy models  we are in luck: the equilibrium distribution is  
confined in a strip, i.e. at least the imaginary part remains bounded. An 
example \cite{Aarts:2013uza} is
\be
\cM=\bR,\; \cM_c=\C,\; S= \frac{1}{2}(1+iB) x^2+\frac{1}{4}x^4\,.
\label{quartic}
\ee

\begin{center} 
\begin{figure}[ht]
\includegraphics[width=0.8\textwidth]{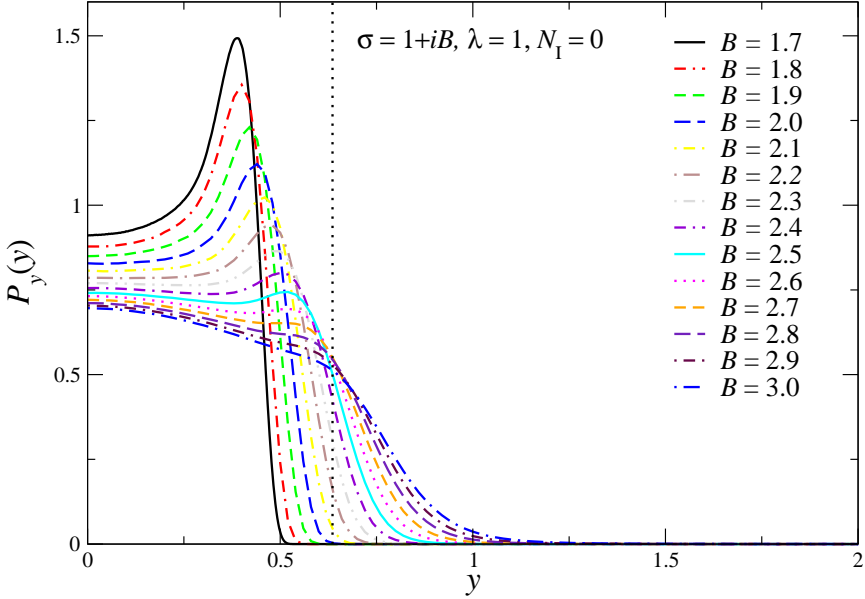}
\caption{Histograms of $P_y(y)$ for various values of $B$ for the toy 
model Eq.(\ref{quartic}).}
\label{quarticplot}
\end{figure}
\end{center}

In this case the equilibrium distribution is confined in a strip provided
$B<\sqrt{3}$ and the CL simulation gives correct results.

The plot Fig.\ref{quarticplot}, taken from \cite{Aarts:2013uza} shows the 
emergence of skirts as $B$ is increasing, crossing the critical value 
$\sqrt{3}$; it displays histograms of the partially integrated equilibrium 
density 
\be 
P_y(y)\equiv\int P(x,y) dx\,. 
\ee

The tails behave like $O((x^2+y^2)^{-3})$ and the results deteriorate as 
$B$ becomes $>\sqrt{3}$.

\section{Boundary terms at poles}\label{sec-12}

This section follows largely our paper \cite{Aarts:2017vrv}.
If $\rho$ has zeroes in $\cM_c$, $\vec K$ has poles there. The evolution 
$\dot{\cO}= L_c\cO $ then generically produces essential singularities at 
these locations.

In equilibrium poles may be
\begin{itemize} 
\item 
outside of the distribution: in this case they are harmless
\item
at the edge of the distribution: the behavior of $|\vec KP|$ determines 
success or failure
\item
inside the distribution: success of failure depend on the behavior of 
$|\vec KP|$ and {\em ergodicity} 
\end{itemize}

Nagata, Nishimura, Shimasaki \cite{Nagata:2016vkn} propose to monitor 
skirts of $|\vec KP|$ at poles as well as at $\infty$. This is an 
excellent idea, allowing to analyze both kinds of boundary terms the same 
way.

\begin{figure}[t]
\centerline{\includegraphics[width=.7\columnwidth]{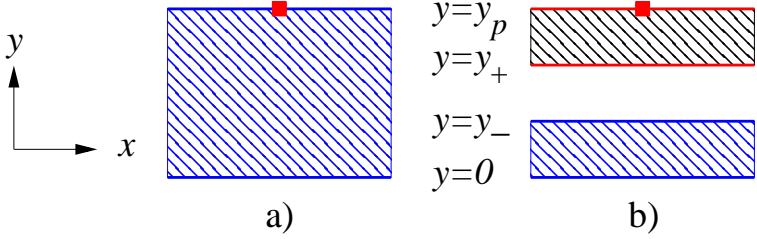}}
\caption{One-pole model Eq.(\ref{rhoonepole}). Left: $\beta<2n_p/y_p^2$, 
right: $\beta>2n_p/y_p^2$.}
\label{strip}
\end{figure}

We study these issues first in a simple one-pole model given by
\be
\rho(x)= (x-iy_p)^{n_p} \exp(-\beta x^2)\,, 
\label{rhoonepole}
\ee
choosing $y_p=1$ for concreteness.
The situation is illustrated in Fig.\ref{strip}: For $\beta < 2n_p/y_p^2$ 
we 
have $P(x,y)> 0$ for $0<y<y_p$ and the pole is on the edge of the 
equilibrium distribution, whereas for $\beta > 2n_p/y_p^2$, $P(x,y)> 0$ if 
and only if $0<y<y_-$ and the pole is outside the equilibrium 
distribution; the upper strip is transient.

\begin{figure}[t]
\centerline{
\includegraphics[width=.4\columnwidth]{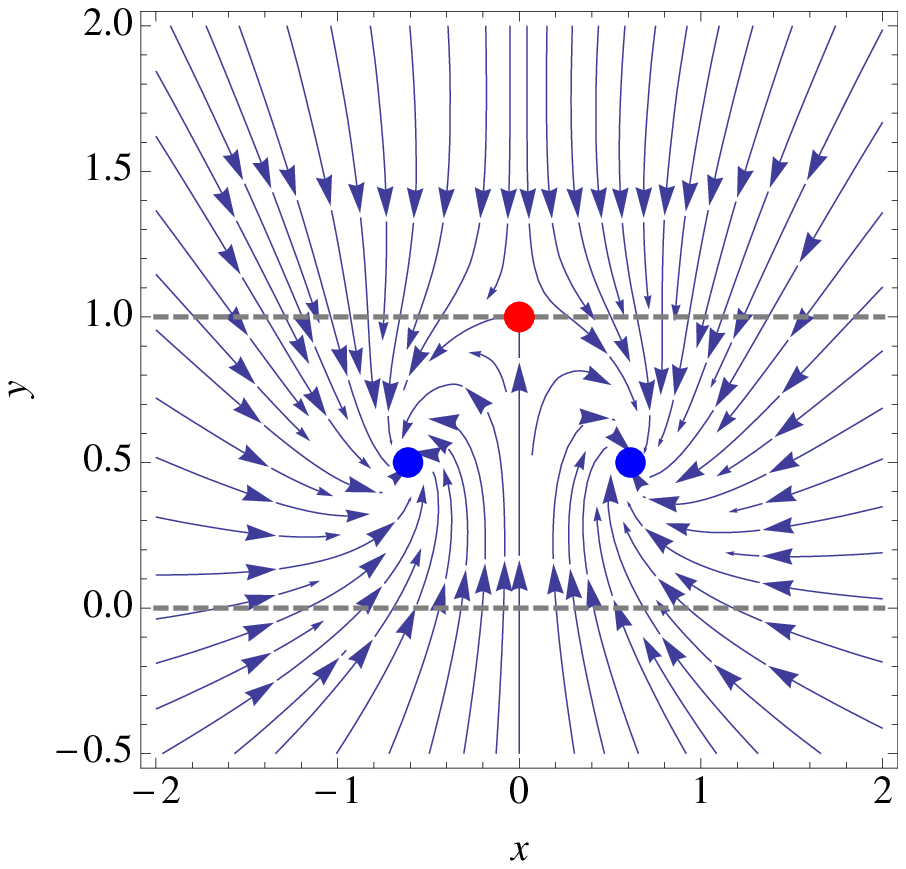}
\includegraphics[width=.4\columnwidth]{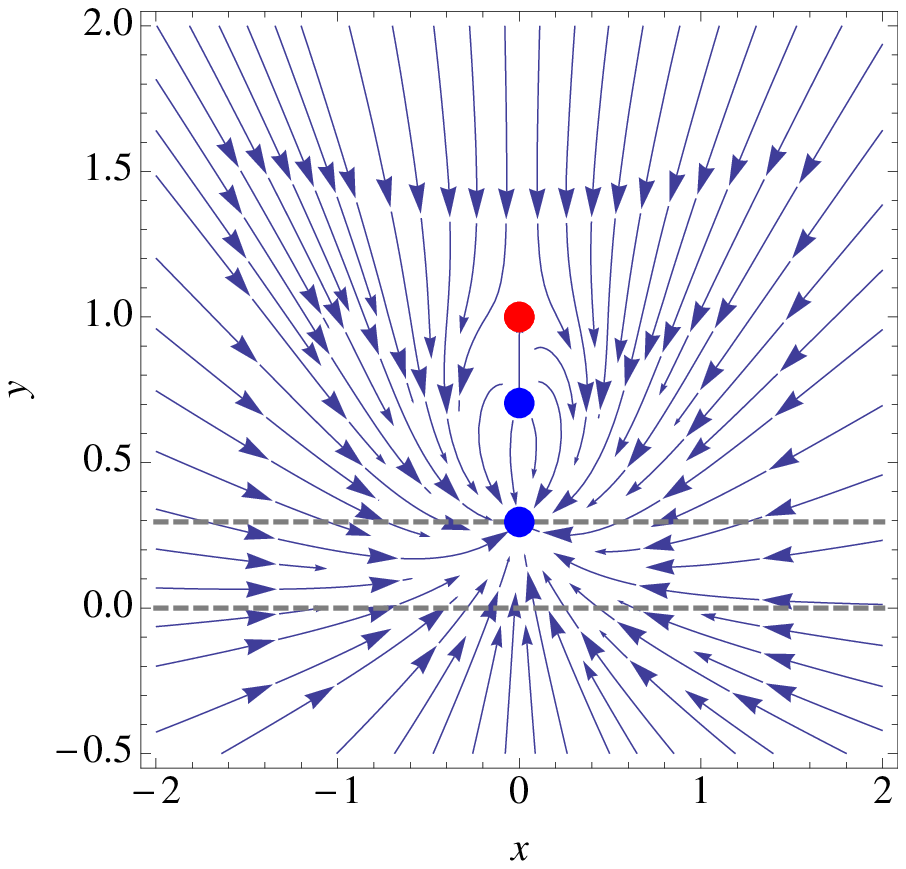}}
\label{onepoleflow}
\caption{`Classical' flow diagram for the model Eq.(\ref{rhoonepole}) with 
$y_p=1,n_p=2$. Blue (red) circles are fixed points (poles).
Left: $\beta=1.6<2n_p$; right: $\beta=4.8>2n_p$; the 
dashed lines indicate the strips confining the equilibrium distribution.}
\end{figure}

\begin{center}   
\begin{figure}[t]
\includegraphics[width=.45\textwidth]{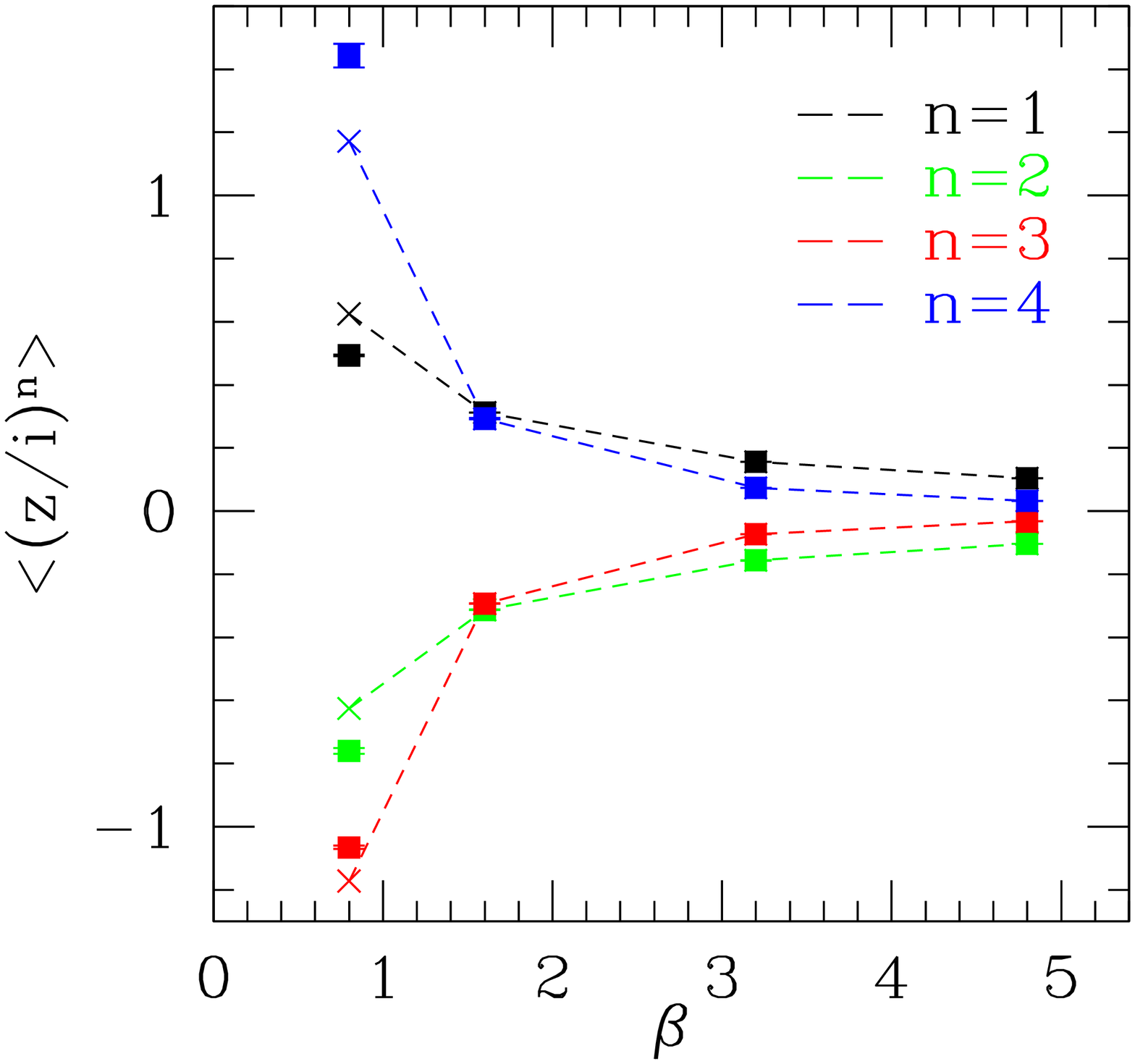}
\includegraphics[width=.45\textwidth]{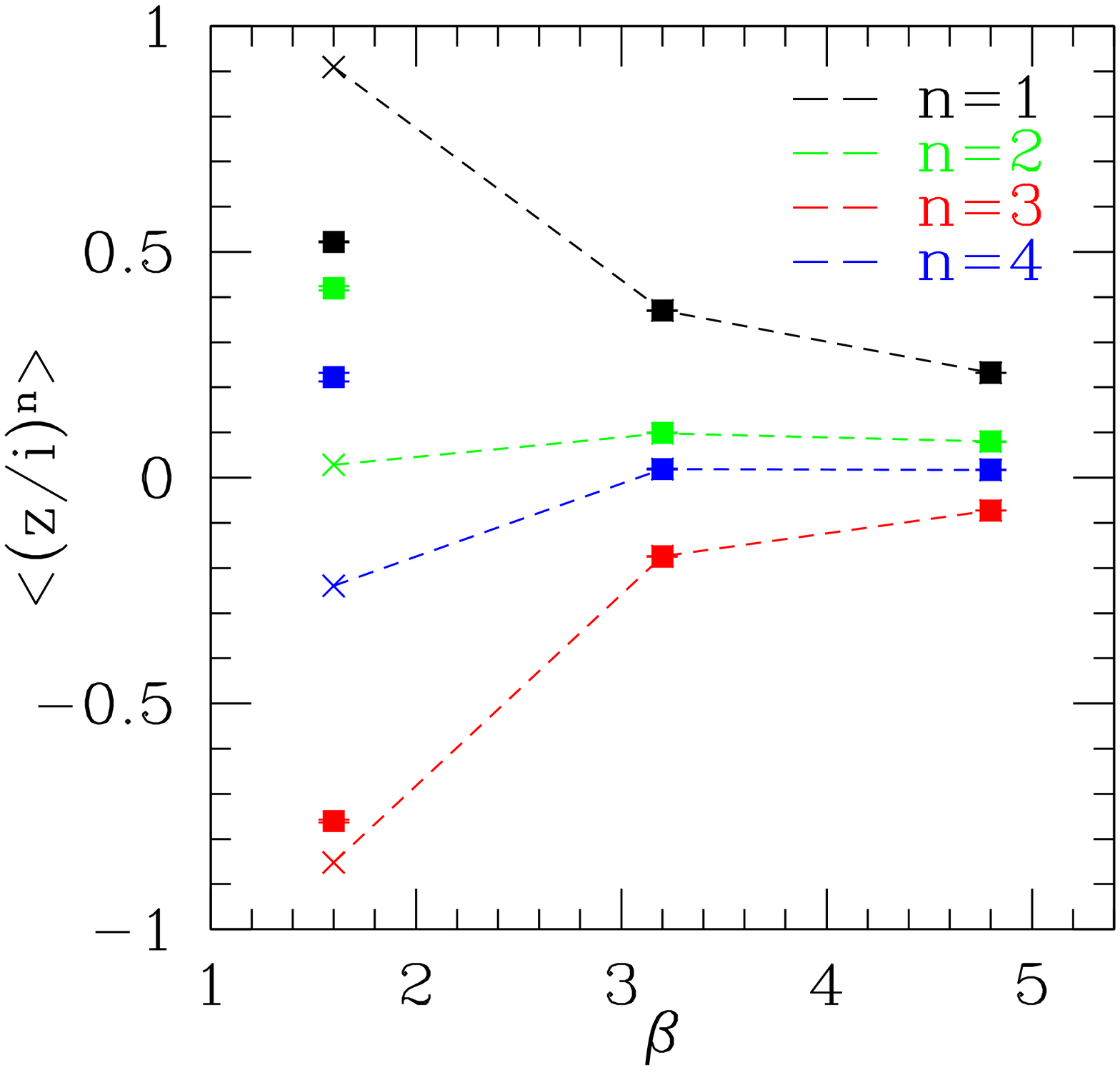}
\caption{Comparison of CL results for $\bra (z/i)^n\ket$ with exact 
results (shown as crosses connected by dashed lines) for the one-pole 
model Eq.(\ref{rhoonepole}). Left panel: $n_p=1$ Right panel:  
$n_p=2$;  $y_p=1$. Errors are smaller than the symbols.}
\label{onepoleresults}     
\end{figure}
\end{center}

\begin{center}
\begin{figure}[t]
\includegraphics[width=.8\textwidth]{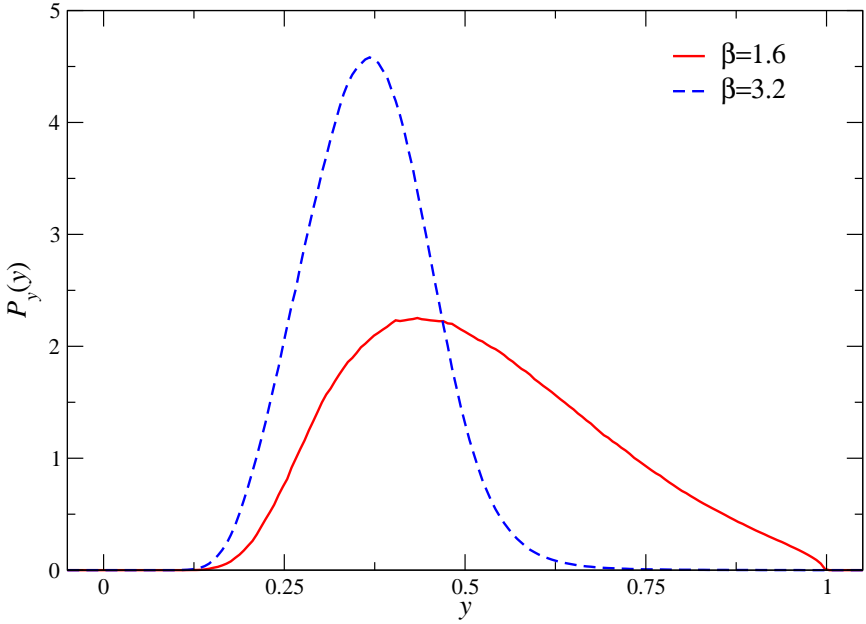}
\caption{Partially integrated distributions $P_y(y)=\int dx P(x,y)$ for 
the one-pole model Eq.(\ref{rhoonepole}).}
\label{py}
\end{figure}
\end{center}

This behavior can be understood by looking at the ``classical'' flow
diagrams, see fig. \ref{onepoleflow} and it can be shown rigorously to be
as stated \cite{Aarts:2013uza,Aarts:2017vrv}.
   
Fig \ref{onepoleresults} shows some results of CL simulations for $n_p=1$
and $\beta=0.8\;,1.6\;,3.2$ and $4.8$ as well as for $n_p=2$ and
$\beta=1.6\;,3.2$. In both cases the results improve dramatically with
increasing $\beta$; for  $n_p=1$, $\beta=1$ is already large enough,
whereas for $n_p=2$ CL results are still quite wrong for this $\beta$.   
The reason for this difference is that $n_p$ is multiplying the pole term
in the drift and so with increasing $n_p$ the pull towards the pole
becomes stronger.

Let us now focus on $n_p=2$.  For $\beta=3.2$ and $4.8$ there is excellent
agreement between simulation results and the exact values. For $\beta=4.8$
(pole outside the distribution) this is to be expected, but it may be
surprising that for $\beta=3.2$, with the pole on the edge, CL still
produces excellent results, unlike the situation for $\beta=1.6$. The 
reason
is the different behavior of the distribution near the poles for
$\beta=3.2$ compared to $\beta=1.6$, as shown in Fig.\ref{py}. For
$\beta=1.6$ there is a considerable boundary term, whereas for $\beta=3.2$
such a boundary term is either absent or invisibly small.

\begin{center}
\begin{figure}[t]
\includegraphics[width=.3\columnwidth]
{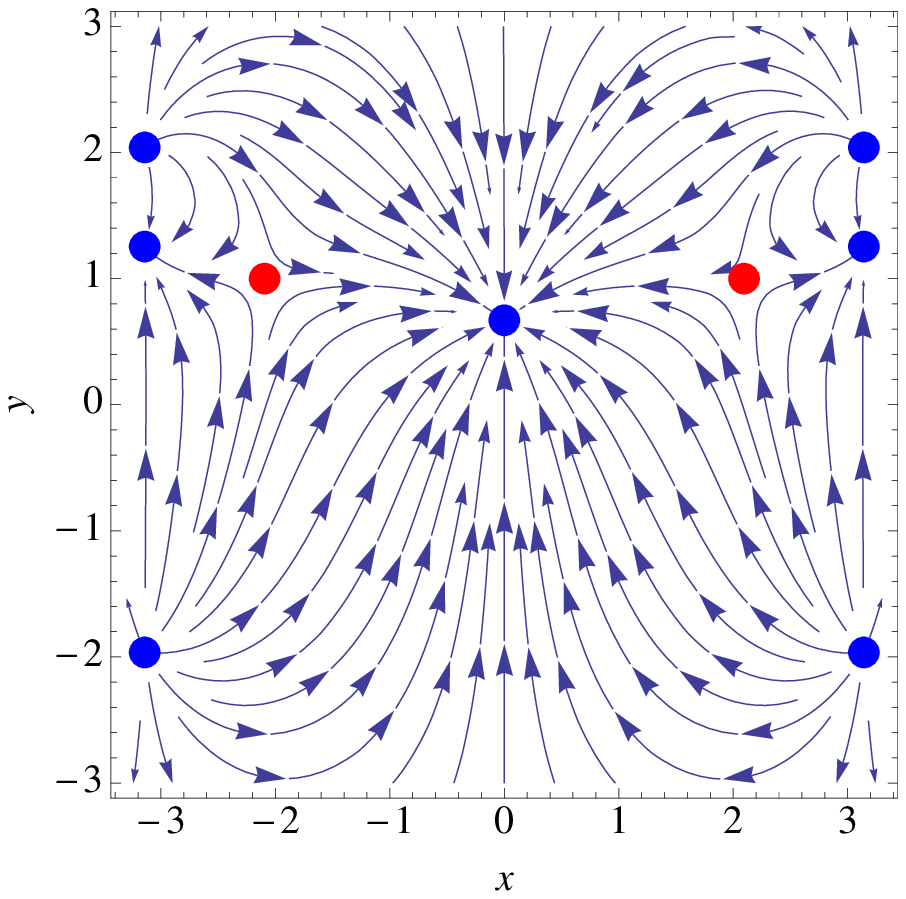}
\includegraphics[width=.3\columnwidth]
{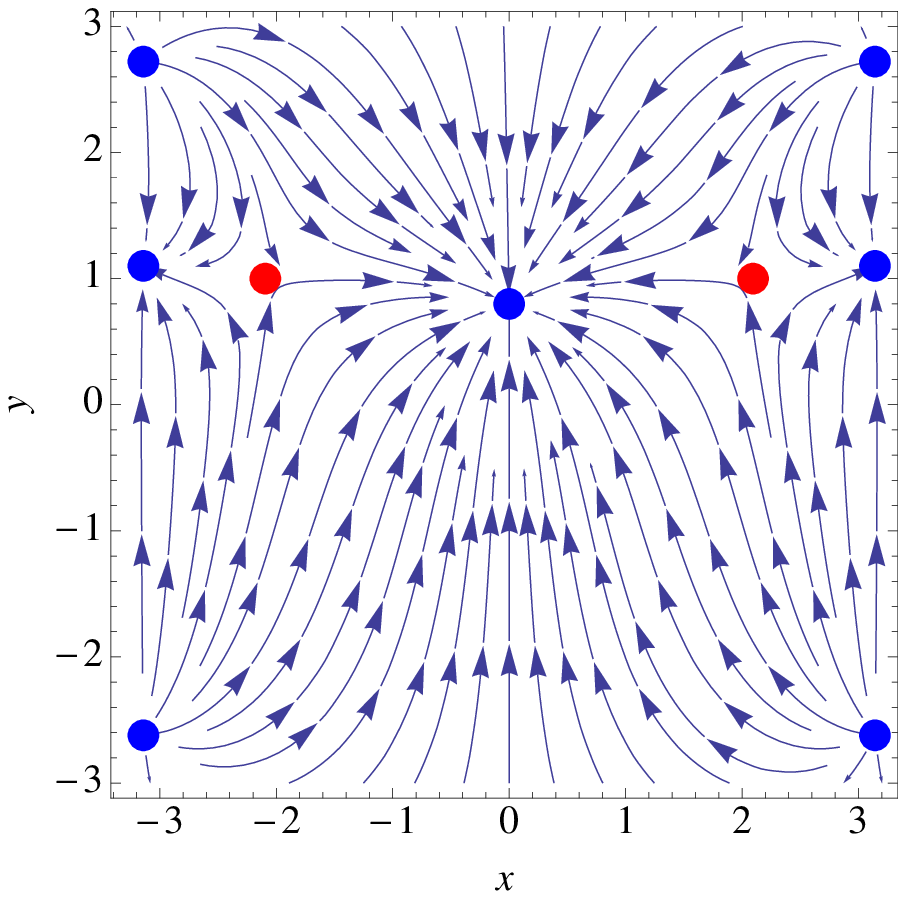}
\includegraphics[width=.3\columnwidth]
{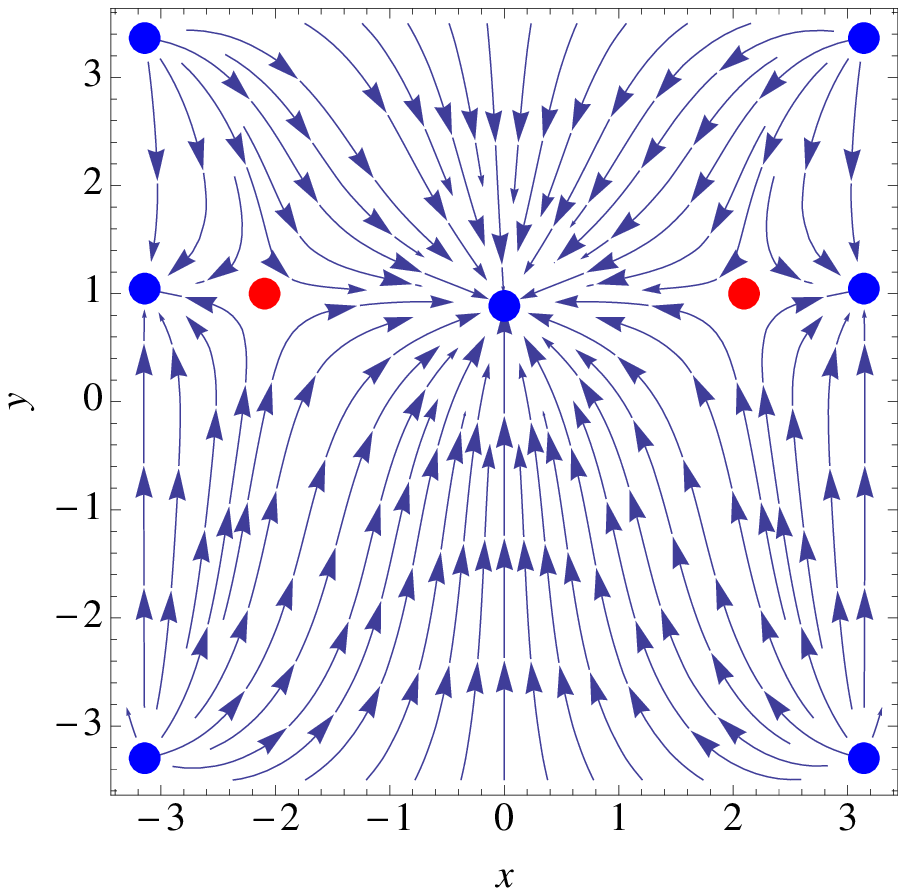}
\caption{Flow portrait for the model Eq.(\ref{onelink}).
Blue dots: fixed points, red dots: poles. Left:  $n_p=1$, 
middle: $n_p=2$, right: $n_p=4$.}
\label{u1flow}
\end{figure}
\end{center}

To see what happens with poles {\em inside} the distribution, we turn to 
another simple model: the one-link U(1) model given by:
\be
\rho(x) = D(x)\exp[\beta \cos(x)];\;\; \kappa=2,\; \mu=1\,,
\label{onelink}
\ee
where
\be
D(x)\equiv \left[1+\kappa \cos(x-i\mu)\right]^{n_p}
\ee
is a mockup of the fermion determinant in a lattice model. We show the 
classical flow portrait for $\beta=0.3$ and three values of $n_p$ in 
Eq.(\ref{u1flow}). Note that besides the attractive fixed point of the 
flow at 
${\Im}\,z=0$ there is a secondary one at ${\Im}\,z=\pm \pi$.

\begin{figure}[t]
\includegraphics[width=.3\columnwidth]
{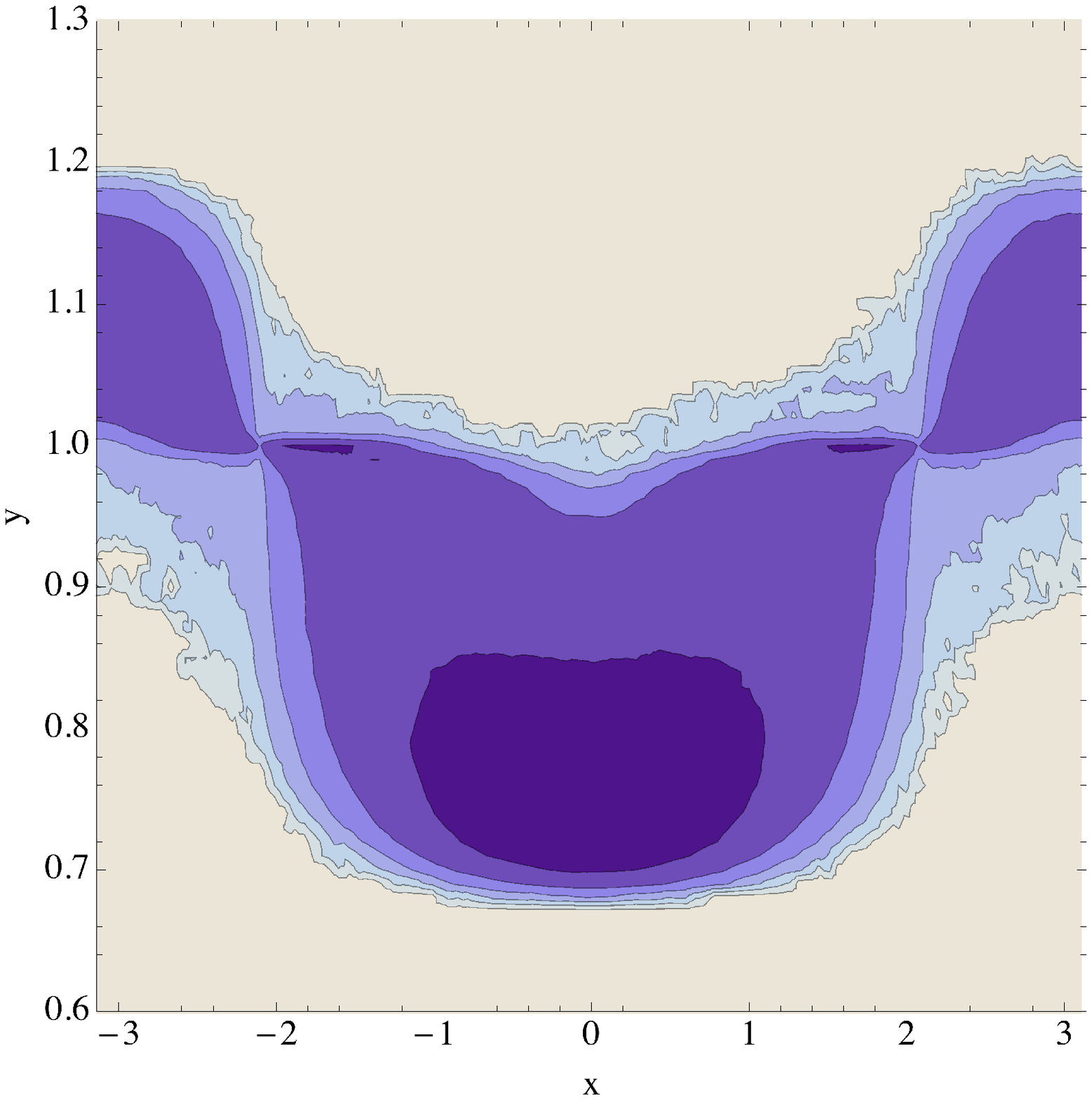}
\includegraphics[width=.3\columnwidth]
{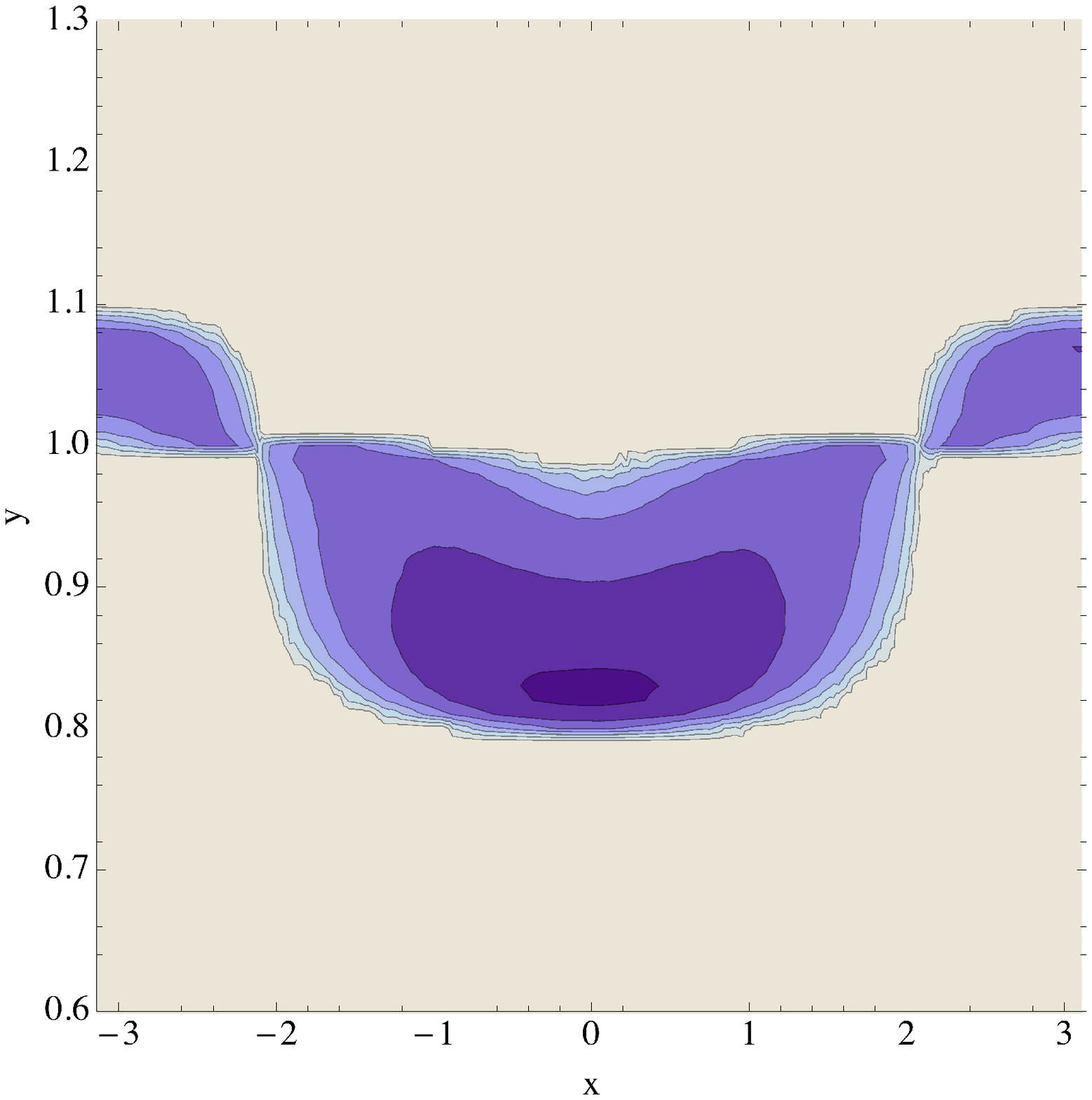}
\includegraphics[width=.3\columnwidth]
{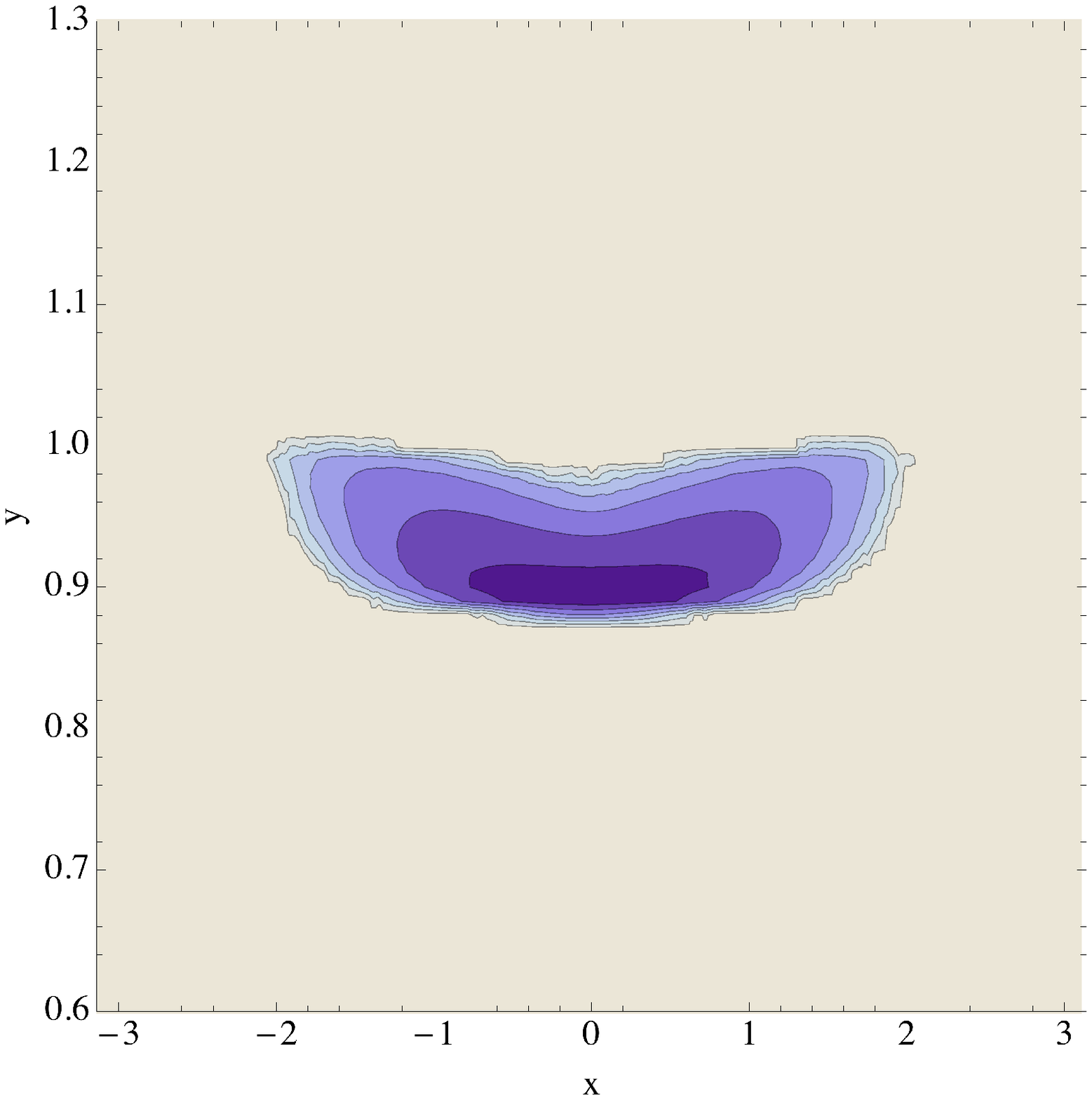}
\caption{Logarithmic contour plots of the equilibrium distributions for 
the model Eq.(\ref{onelink}). Left:  $n_p=1$, middle: $n_p=2$, right: 
$n_p=4$.}
\label{equilu1}
\end{figure}

The equilibrium distributions for the same cases are shown in 
Fig.\ref{equilu1}. 
For $n_p=1,2$ they show two almost separated regions defined by
\be
G_\pm\equiv\{z\in\C|\,{\rm sgn}\, {\Re}\, D(z)=\pm 1\}\,
\ee
visible as the `head' and `ears', respectively. For $n_p=4$ we only see 
one region contained in $G_+$.

It can be seen that the process tends to avoid the poles, creating 
bottlenecks there. But it does not produce correct results (see 
\cite{Aarts:2017vrv}). The reason will be discussed in the next section, 
where we find that the CL processes restricted to $G_+$ or $G_-$  
actually simulate correctly a {\em different complex measure}.

\begin{center}
\begin{figure}[t]
\includegraphics[width=.45\textwidth]{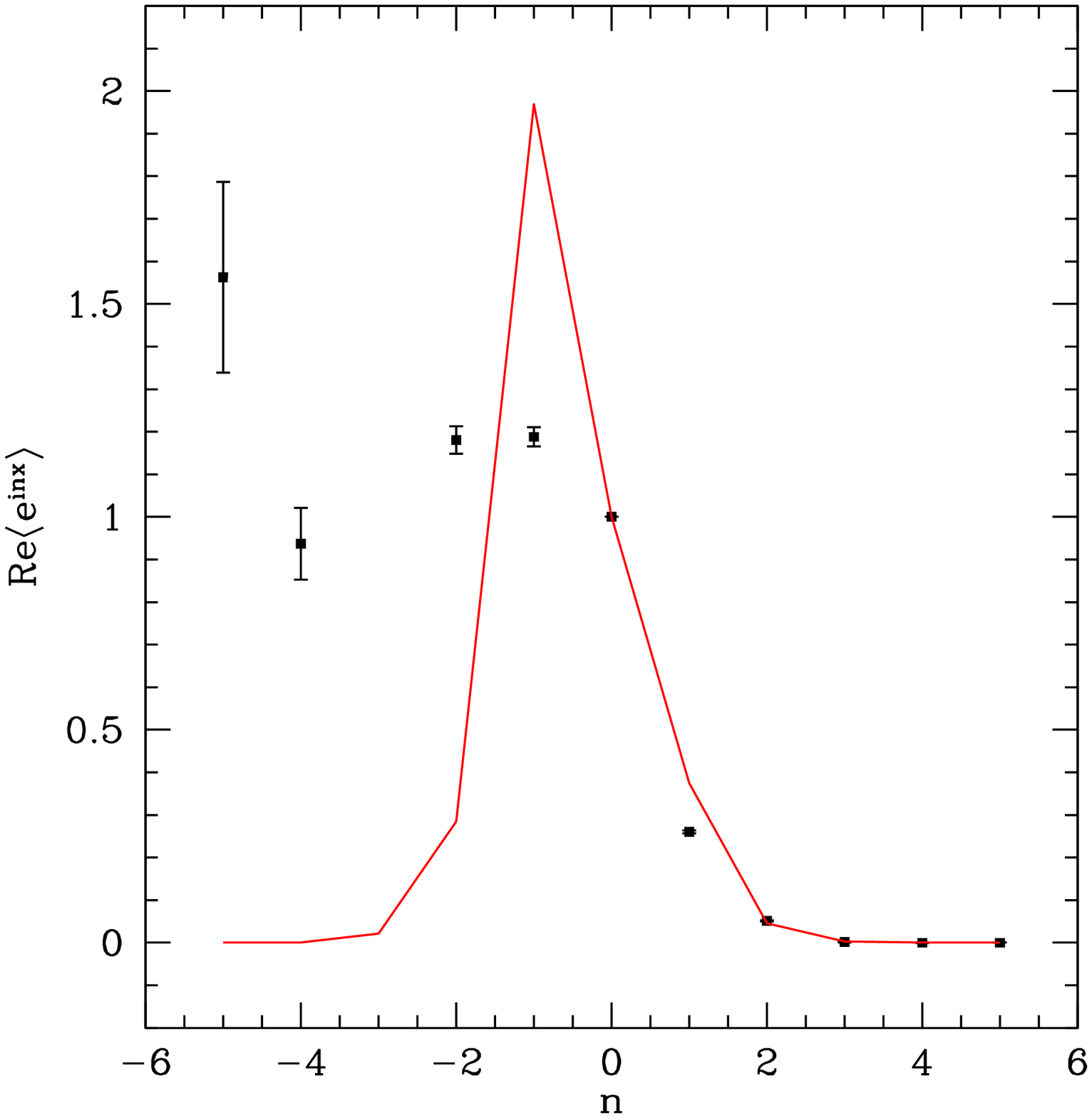}
\includegraphics[width=.45\textwidth]{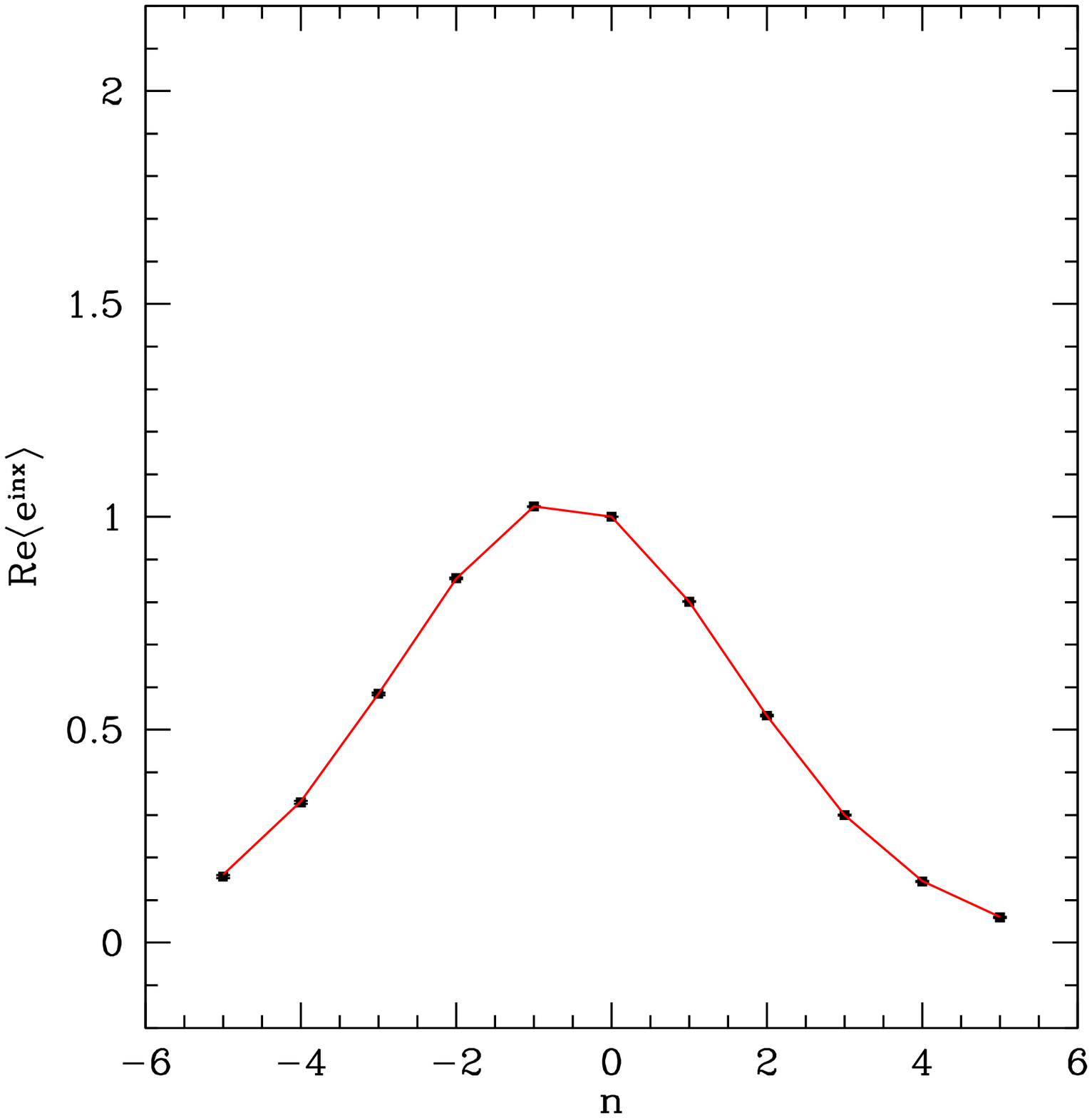}
\caption{Comparison of CL simulation results with exact values for the 
$U(1)$ one-link model. Shown are modes $\bra\e^{inz}\ket$ for 
$n=-5,\ldots,5$.  Left: $\beta=0.3$, right: $\beta=5$.}
\label{betacomp}
\end{figure}
\end{center}

It is also interesting to study the effect of $\beta$ here as well. We 
compare the results of simulations for $\beta=0.3$ and $\beta=5$ with the 
exact results in Fig.\ref{betacomp} and see the failure of the simulation 
for 
$\beta=0.3$, whereas for $\beta=5$ there is perfect agreement, even though 
in both cases the drift has a pole in the same place. This of course due 
to the fact that the large value of $\beta$ makes the attraction of the 
fixed point at $x=0$ very strong.

\section{Failure of ergodicity}

Consider a special case of the real one-pole model:
\be
\rho=x^4 \exp\left(-\frac{x^2}{2\sigma} \right)\,;\quad
K=-\frac{x}{\sigma}+\frac{4}{x}\,.
\ee
The Fokker-Planck Hamiltonian related to this model (see Section \ref{sec-2}) 
is
\be
H_{FP}=-\frac{d^2}{dx^2}+\frac{2}{x^2}+\frac{x^2}{2\sigma}-
\frac{5}{2\sigma}\,;
\ee
$H_{FP}$ has {\em} two ground states 
\be
\Omega_\pm\propto \theta(\pm x) x^2  \exp\left(-\frac{x^2}{4\sigma}
\right)\,.
\ee
Corresponding to these ground states are {\em two} equilibrium probability 
densities:
\be
P_\pm(x)\propto \Omega_\pm^2\propto \theta(\pm x) x^4  
\exp\left(-\frac{x^2}{2\sigma}
\right)\,.
\ee
The pole at the origin is a bottleneck, obstructing the crossing of the 
real CL process. 

Returning now to the one-link U(1) model Eq.(\ref{onelink}) 
(Section \ref{sec-5}), 
recall that there the poles are also bottlenecks. We venture a bold 
generalization:
\vglue2mm 
\begin{center} 
{\em \large Poles inside the distribution tend to form bottlenecks!} 
\end{center} 
\vglue2mm

In the simulation for $n_p=1,2$ , the process occasionally manages to 
cross between the regions $G_+$ and $G_-$, but this might be entirely due 
to the nonzero step size. For $n_p=4$ no such crossing was observed, even 
in very long runs (Langevin time 25000).

Restricting the simulation process to $G_+$ ($G_-$), it turns out that it 
correctly simulates the integral from one pole to the other along a path 
$C_+(C_-)$ from one pole two the other contained in $G_+$ ($G_-$)! $C_+$ 
is the path starting at the pole with negative real part and ending at the 
one with positive real part, whereas $C_-$, using the periodicity of the 
system starts at the latter pole and moves right, ending at the first 
pole.

\begin{center}
\begin{figure}[t]
\includegraphics[width=.3\textwidth]{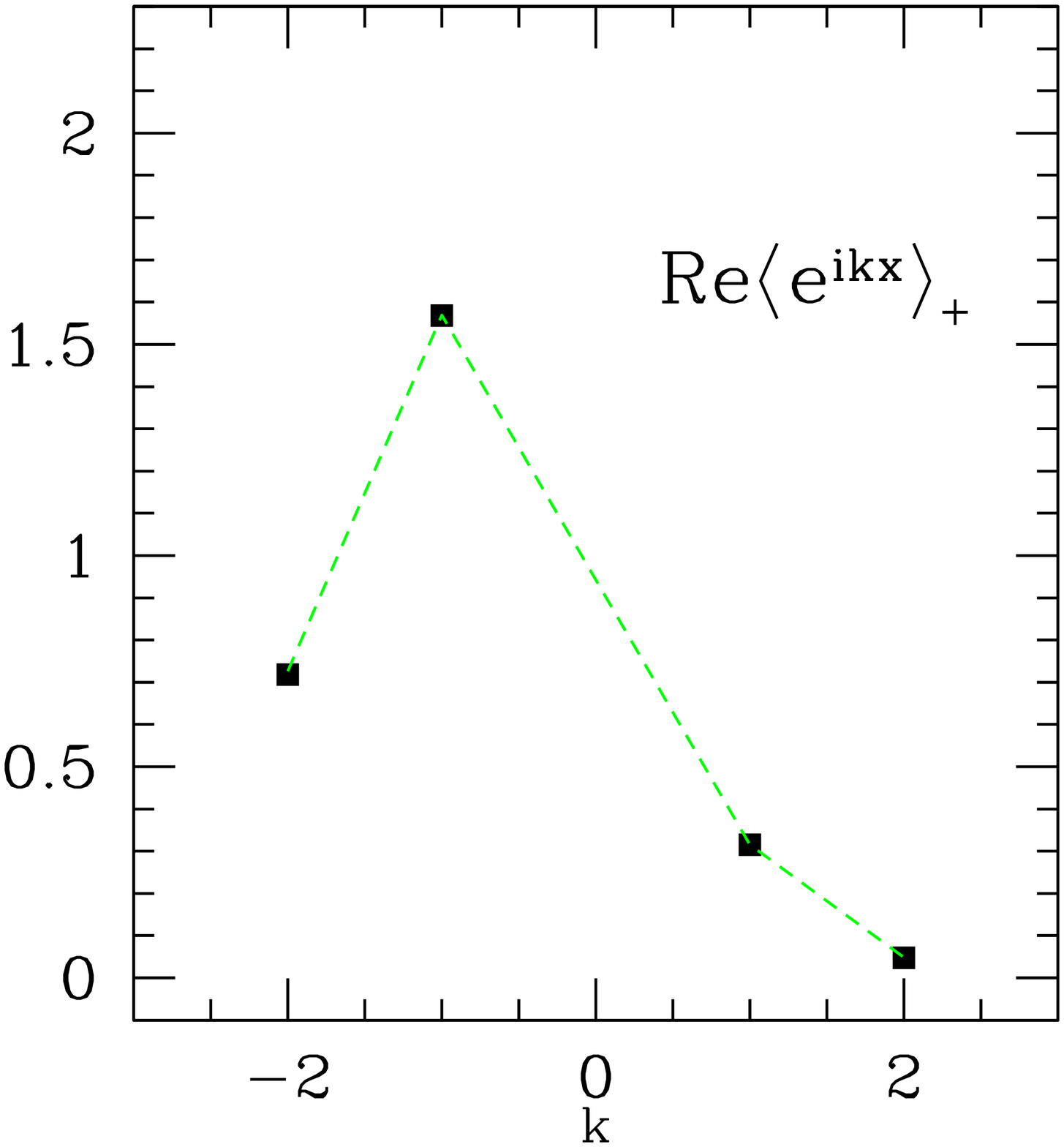}
\includegraphics[width=.3\textwidth]{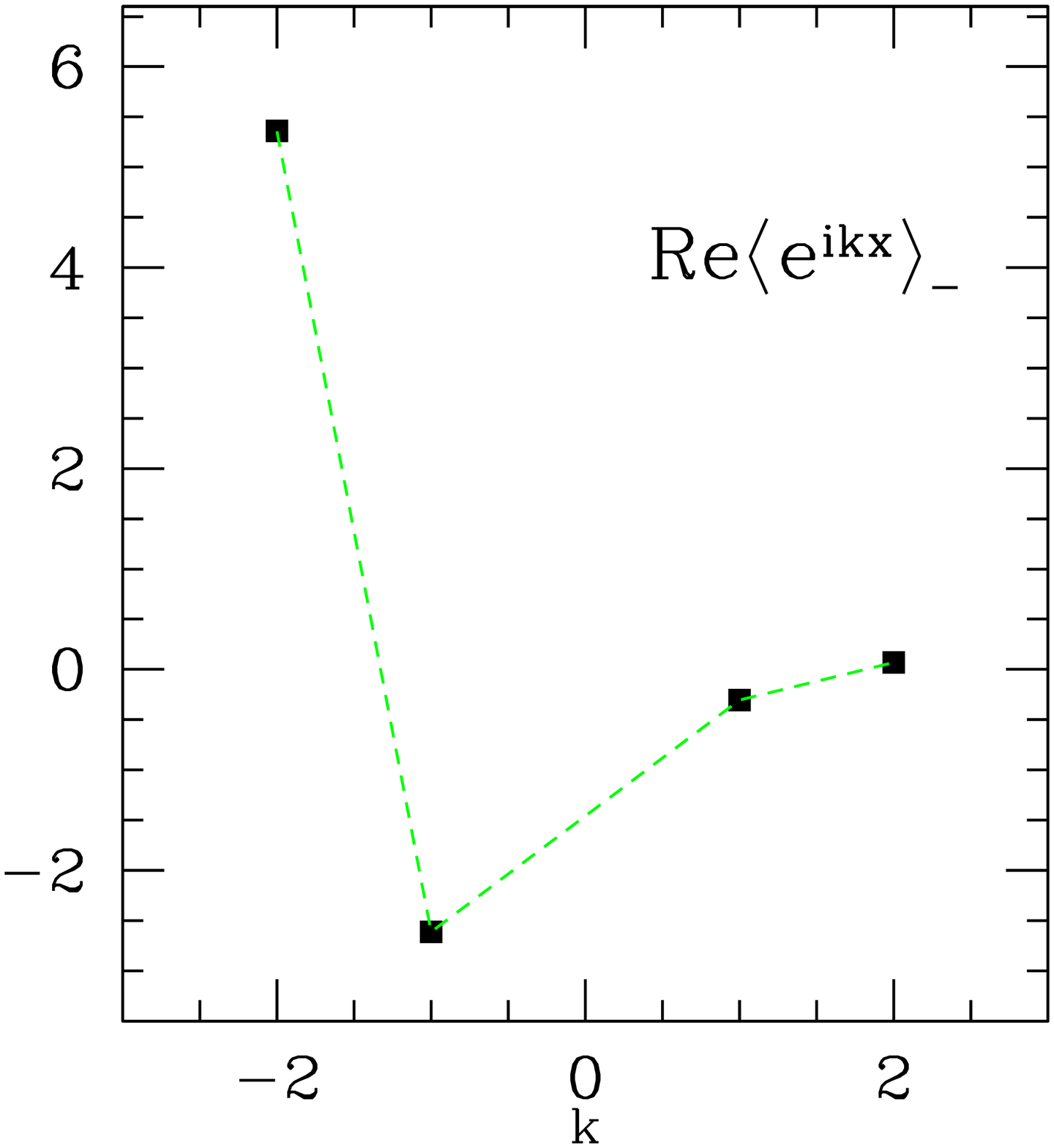}
\includegraphics[width=.3\textwidth]{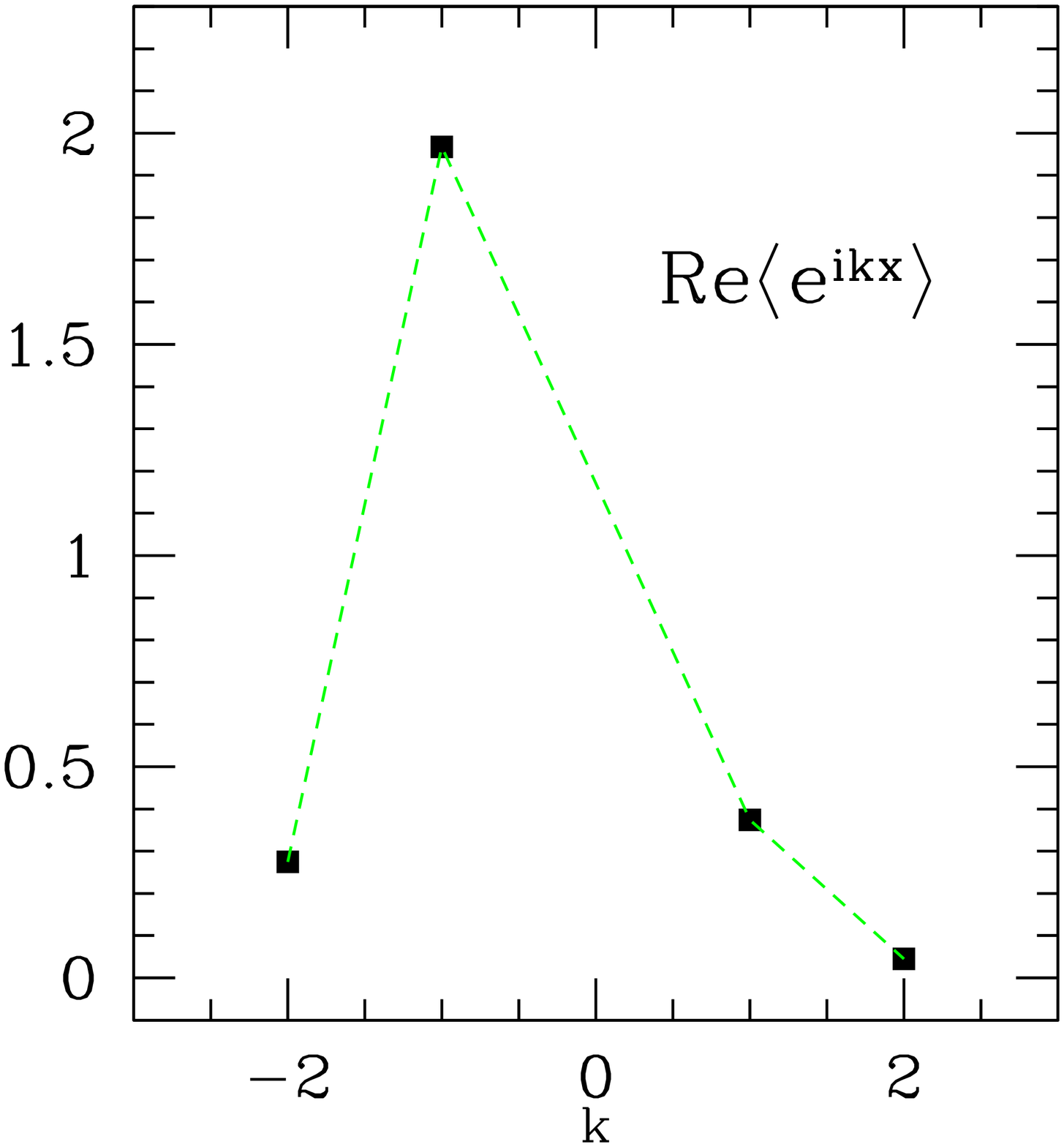}
\caption{Re$\bra e^{ikz}\ket$ for $n_p=1$ and several values of $k$; left:
restricted to $G_+$, middle: restricted to $G_-$, right: combined with
the weights $w_+=1.09551, w_-=-0.09551$. Black dots are results of CL
simulations, green dashed lines connect the exact values; errrors are 
smaller than the symbols.}
\label{restr1}
\end{figure}
\end{center}

\begin{center}
\begin{figure}[t]
\includegraphics[width=.3\textwidth]{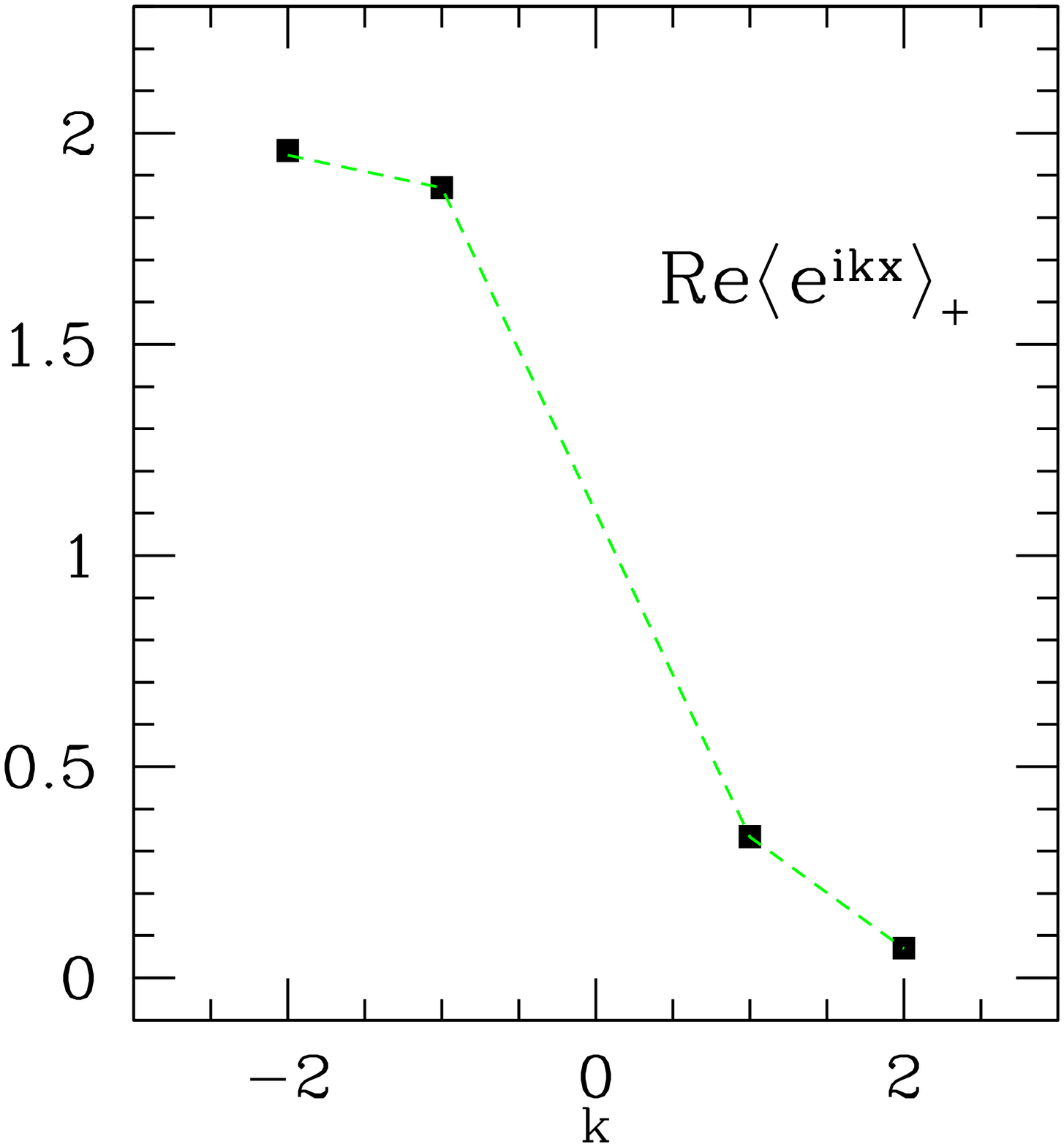}
\includegraphics[width=.3\textwidth]{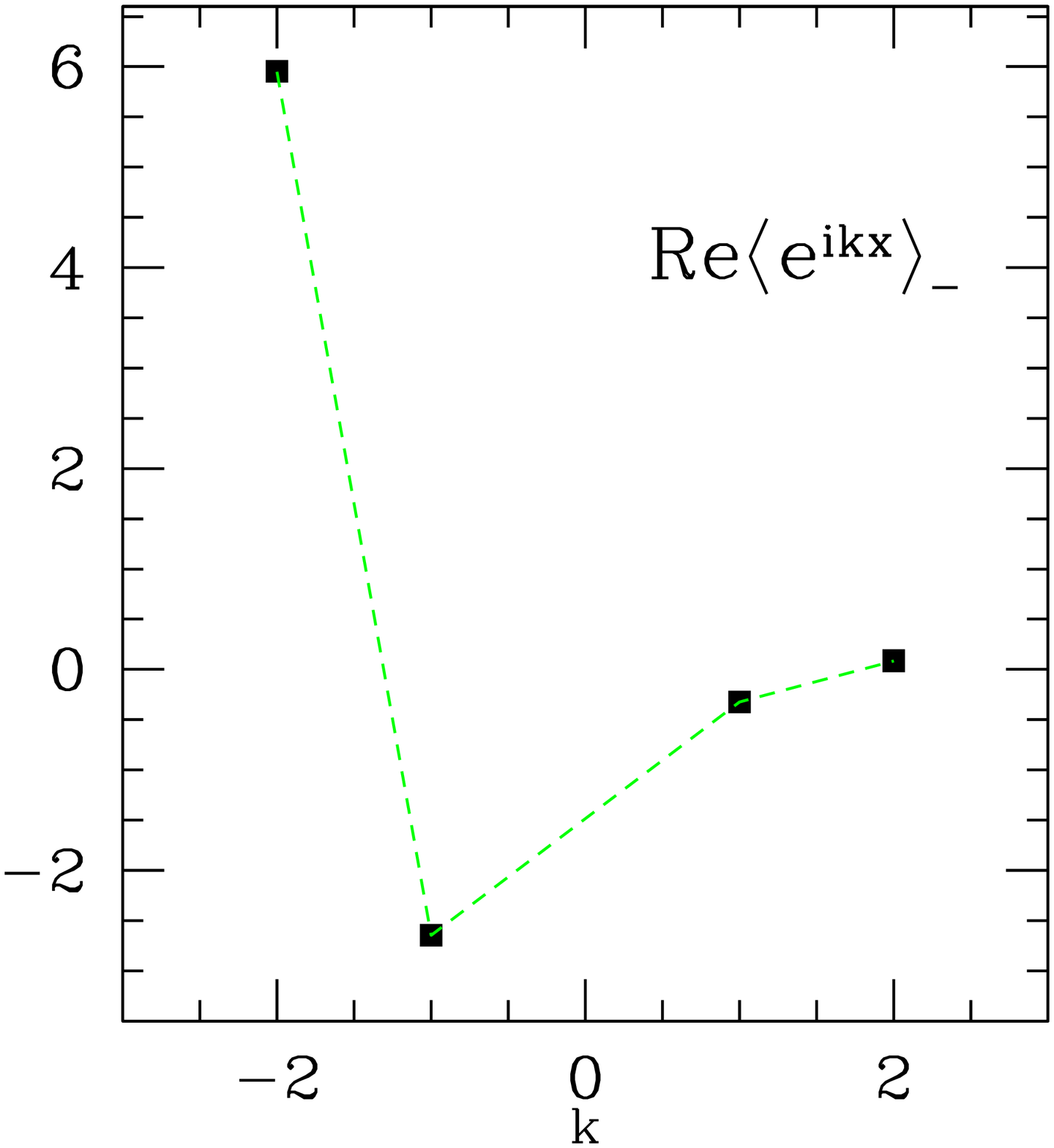}
\includegraphics[width=.3\textwidth]{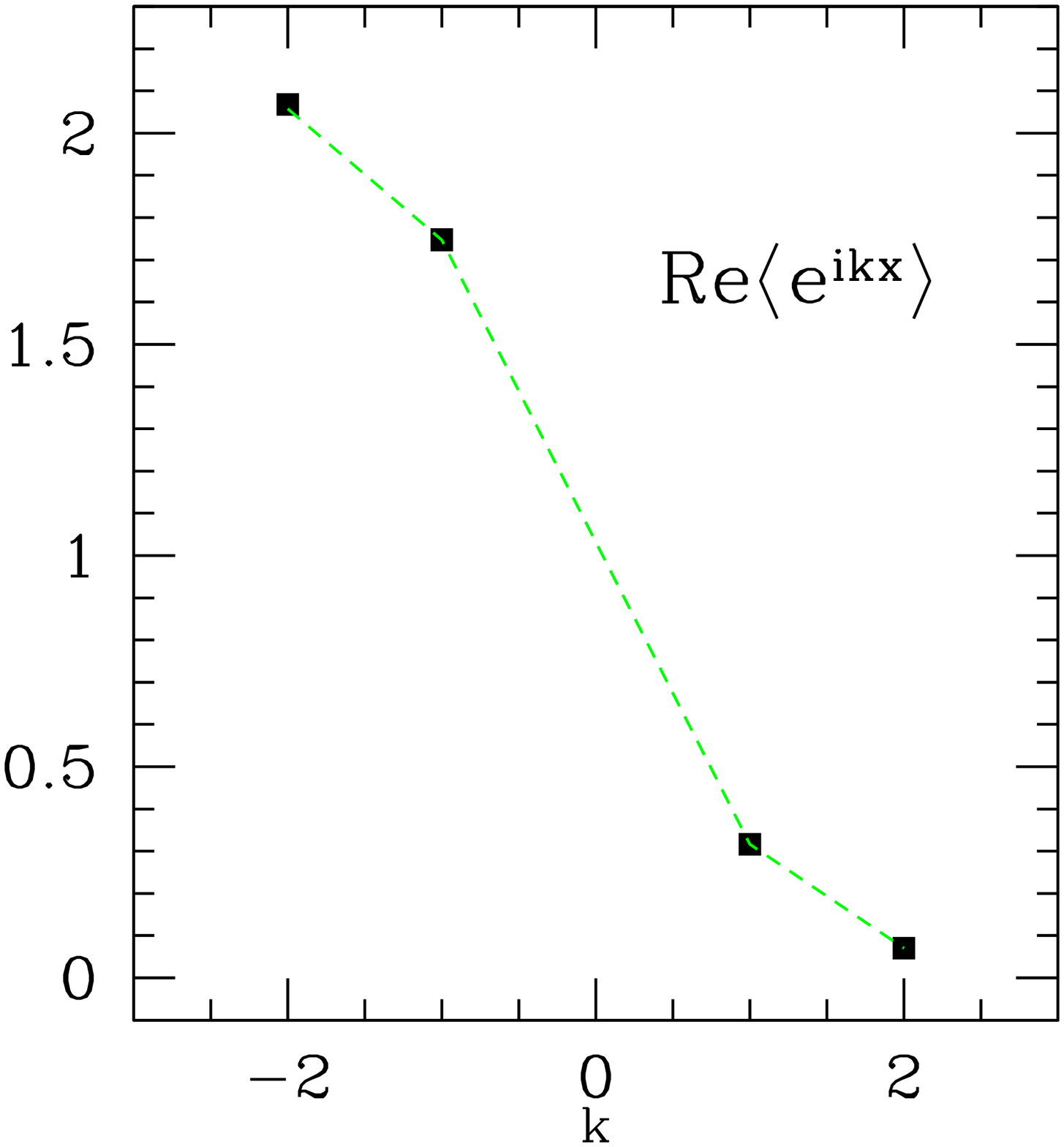}
\caption{Same as Fig.\ref{restr1}, but for $n_p=2$, and 
weights $w_+=0.97267, w_-=0.02733$.}
\label{restr2}
\end{figure}
\end{center}

This is shown in Fig.\ref{restr1} for $n_p=1$ and in Fig.\ref{restr2} for 
$n_p=2$. So probably for $n_p=1,2$ and definitely for $n_p\ge 4$ there are 
two invariant (equilibrium) distributions and the process is {\em 
nonergodic}!

In this simple model one can actually obtain correct results for the 
original problem by combining the two restricted processes with suitable, 
not always positive weights, $w_\pm$:
\be
\bra\cO\ket\equiv w_+ \bra\cO\ket_+ + w_- \bra\cO\ket_-\,,
\ee
where $\bra\cO\ket_\pm$ are results of the CL simulation restricted to 
$G_\pm$.
  
$w_+$ and $w_-$ are easily determined: they are \be 
w_\pm=\frac{Z_\pm}{Z_++Z_-}\,, \ee with \be Z_\pm=\int_{C_\pm} \rho(z) 
dz\,, \ee
For our parameter set $\beta=0.3,\kappa=2, \mu=1$ the weights are: 

\quad $n_p=1:$  $w_+=1.09551,\; w_-=-0.09551 $,

\quad $n_p=2:$  $w_+=0.97267,\; w_-=0.02733 $,

\quad $n_p=4:$  $w_+=0.99699,\; w_-=0.00301 $.

\noindent
(cf \cite{Aarts:2017vrv}). So the `ears' region $G_-$ has already rather 
small weight for $n_p=1$, getting even smaller with increasing $n_p$, so 
that for $n_p=4$ the `head' region $G_+$ alone gives results so close to 
the exact ones as to be practically indistinguishable.

The point to take home from this discussion is: absence of boundary terms, 
i.e. sufficiently fast vanishing of the distribution approaching 
a pole is {\em necessary} for correctness. It is not, however, {\em 
sufficient}, because nonergodicity may mean that a restriction of the 
desired complex measure is being simulated.

\subsection{Poles in QCD}\label{sec-13}

The fermion determinant in QCD 
\be 
\det(\DD\,_U+M)
\ee
in a finite volume is a polynomial in the matrix elements of the direct 
product of the $SL(3,\C)$ groups forming $\cM_c$, so it {\em always} has 
zeroes somewhere in $U\in SL(3,\C)$. These zeroes are of course not 
isolated points, but rather form submanifolds of $\cM_c$ of codimension 
2. It should be expected that this also may lead to boundary 
terms and/or non-ergodicity, which might be the reason for failure in 
some cases.  
\vskip4mm
But there is some good news: Sexty \cite{Sexty:2013ica} and Aarts et al  
\cite{Aarts:2017vrv} find that the eigenvalues avoid $0$, at least for 
the parameters studied, so at least there are apparently no boundary 
terms here.

\begin{figure}[ht]
\begin{center}
\includegraphics[width=0.4\columnwidth]{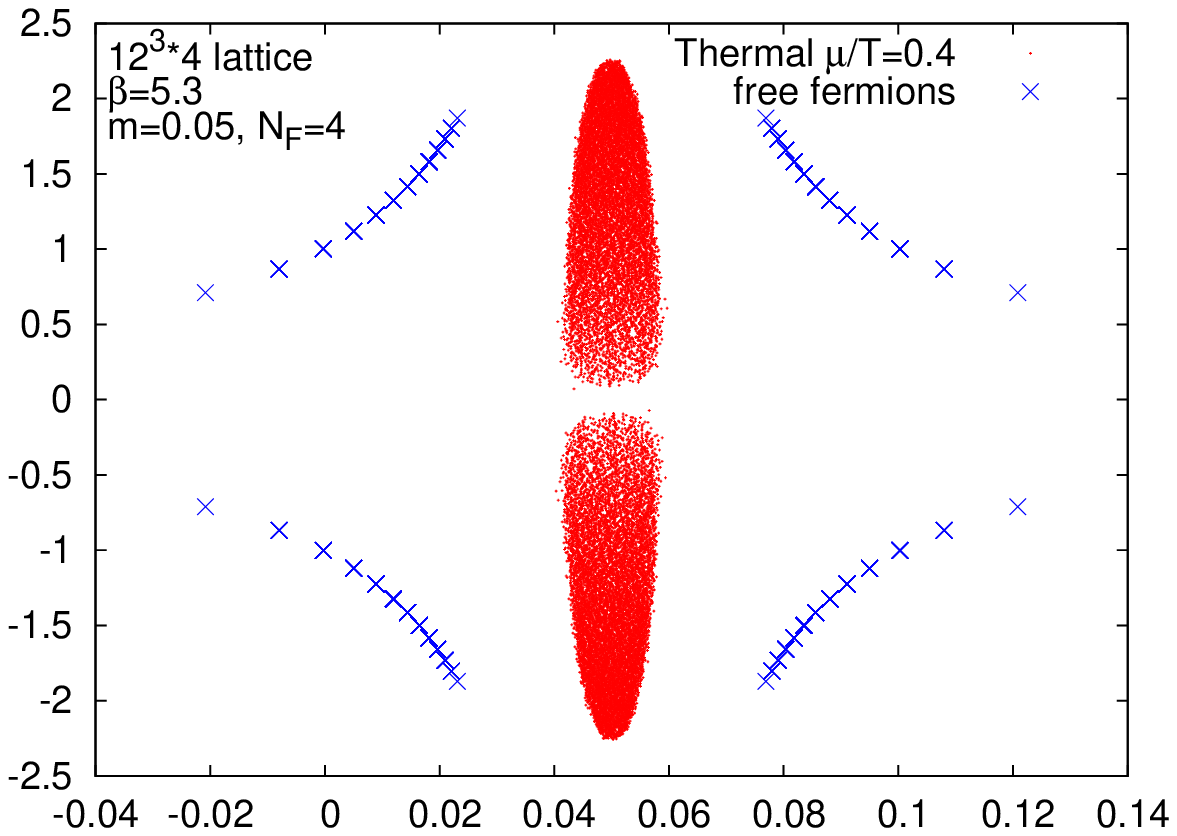}
\includegraphics[width=0.4\columnwidth]{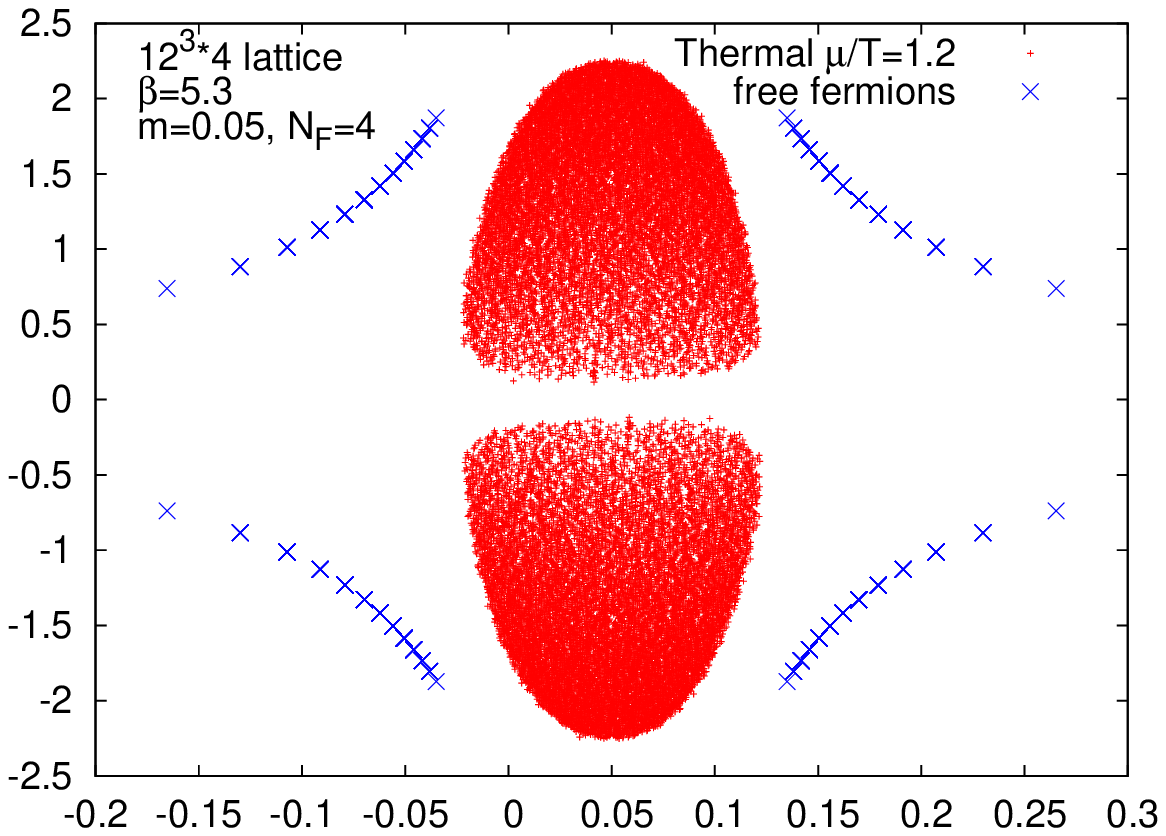}
\includegraphics[width=0.4\columnwidth]{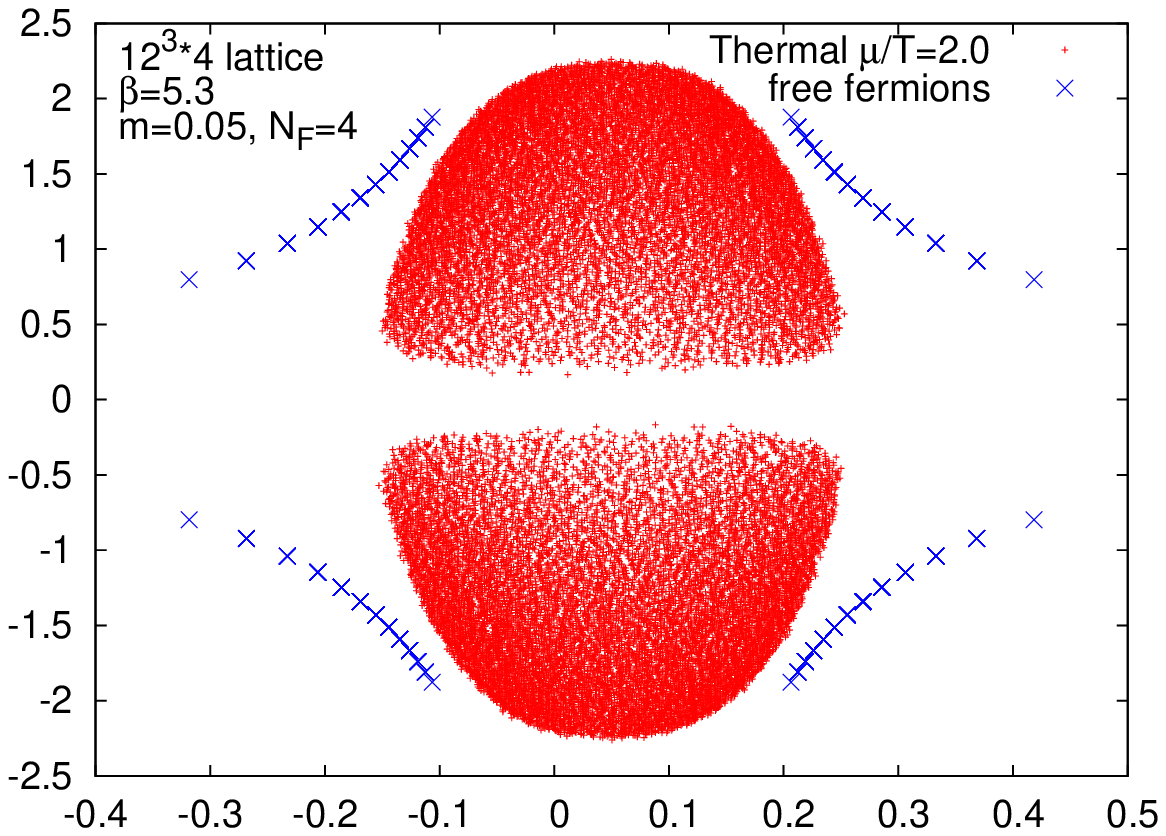}
\includegraphics[width=0.4\columnwidth]{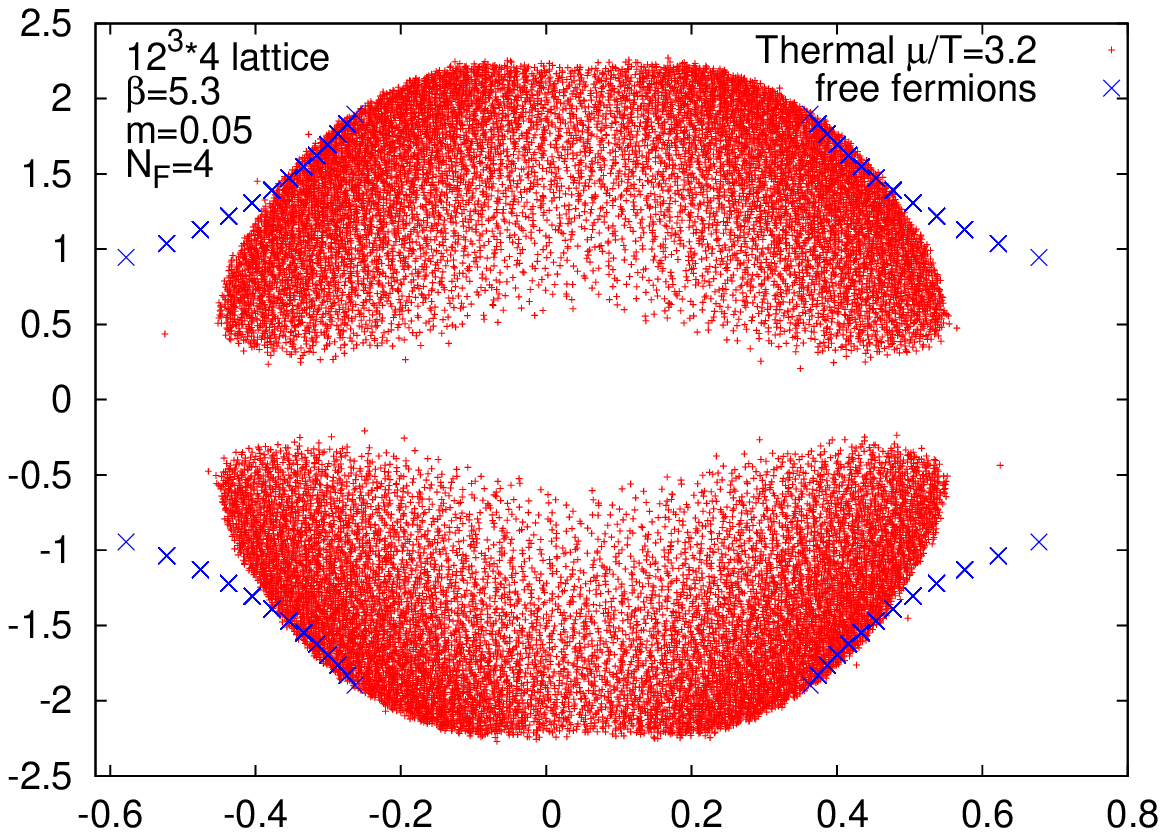}
\label{spectra12}
\caption{Spectrum of staggered Dirac op., $12^3\times 4$ lattice,
$\mu/T=0.4, 1.2, 2.0, 3.2$.}
\end{center}
\end{figure}

It can be hoped that in many cases the situation is as in the one-link 
$U(1)$ model for $n_p=4$, i.e. that the region $G_+$, where $\Re\, 
det(\DD_U+m)\ge0$ already gives results sufficiently close to the 
correct ones. 

\section{Gauge cooling (GC)}\label{sec-14}

Gauge cooling  \cite{Seiler:2012wz} is a necessary (not always sufficient) 
procedure for obtaining stable results from CL simulations of lattice 
gauge theories. It starts with the observation that under complexification
of the configuration space of a lattice gauge theory the group of gauge 
transformations gets complexified as well. 

Let's be concrete and focus on lattice QCD. After integrating out the 
fermions the configuration space is $\cM\equiv SU(3)^{\times N_l}$ which 
gets complexified to $\cM_c\equiv SL(3,\C)^{\times N_l}$, where $N_l\equiv 
\#$ of links.  The gauge group $\cG\equiv SU(3)^{\times N_s}$ is 
complexified to $\cG_c\equiv SL(3,\C)^{\times N_s}$ with $N_s\equiv \#$ of 
sites. The point is that observables $\cO$, analytically continued to 
$\cM_c$ and invariant under $\cG$ are also invariant under the larger,  
noncompact group $\cG_c$. So in the ideal world of mathematics the 
expectation values of $\cO$ will not change under gauge transformations 
from $\cG_c$; they also would not change if the Langevin process is 
modified by adding a component tangential to the gauge orbit to the drift 
$\vec K$.

Nevertheless, such a modification of the process is necessary for the 
following reason: the Langevin process will move, due to buildup of 
rounding errors, exponentially fast into noncompact directions, far away 
from the unitary submanifold $\cM$, because the drift $\vec K$ produced 
from a gauge invariant action is gauge covariant and has no component 
restricting the movement along the gauge orbits of $\cG_c$. The link 
variables thus will soon be very large, whereas the gauge invariant 
observables composed of them will involve huge cancellations, so that 
their values will become unreliable.  The process will become numerically 
uncontrolled. This exponential growth is seen in Fig.\ref{unitnorm}, taken 
from \cite{Seiler:2012wz} (which will be explained later). So some 
stabilization of the noncompact directions is definitely necessary.

This can be done either by interspersing some `gauge cooling steps' 
between `dynamical updates' \cite{Seiler:2012wz} or by adding a suitable 
cooling term to the drift, as advocated by \cite{Nagata:2016alq}. A 
cooling procedure first requires defining a suitable distance from the 
unitary submanifold $\cM$, called `unitarity norm'; a simple choice is
\be
F(\{U\})\equiv \sum_{\vec x}\tr\left[\, {U_{\vec x}}^\dagger U_{\vec x}
+({U_{\vec x}}^\dagger)^{-1}{U_{\vec x}}^{-1}-2\right]\ge 0\,,
\ee
where the sum is over the lattice sites; but of course there are other 
possibilities (see for instance \cite{Aarts:2014kja,Nagata:2016vkn}, 
which are sometimes preferable). $F(\{U\})$ vanishes if and only if all 
the $U$'s are unitary.

Dynamical updates are defined, using a simple Euler discretization, by
\be
U_{\vec x,\mu}\mapsto \exp\left\{-\sum_a i\lambda_a (\epsilon
K_{a,\vec x,\mu}+\sqrt{\epsilon}\eta_{a,\vec x,\mu})\right\}U_{\vec x,\mu}\,,
\label{dyn}\ee
where the $\lambda_a\,,a=1,\ldots,8$ are the Gell-Mann matrices forming a
basis of the Lie algebra of $SU(3)$ and the drift is given by
\be
K_{a,\vec x,\mu}= D_{a,\vec x,\mu}S
\label{drift}
\ee
with the derivation $D$ acting on a function $f$ as
\be
D_{a,x,\mu} f(\{U\})= \lim_{\delta\to 0}
\frac{1}{\delta}\left [ f(\{U(\delta)\}) -f(\{U\})\right]\,,
\label{deriv}
\ee
where $\{U(\delta)\}$ means the variable $U_{x,\mu}$ has been replaced by
$\exp(i\delta \lambda_a)U_{x,\mu}$ with all other variables unchanged.
We define a `gauge gradient' of the unitarity norm by
\be
G_{a,\vec x}\equiv D_{a,\vec x}F= 2\tr \lambda_a \left[U_{\vec x,\mu}   
U^\dagger_{\vec x,\mu}-
U_{\vec x-\hat\mu,\mu}^\dagger U_{\vec x-\hat\mu,\mu}\right]
+2 \tr \lambda_a \left[-(U_{\vec x,\mu}^\dagger)^{-1}U_{\vec x,\mu}^{-1}
+(U_{\vec x-\hat\mu,\mu}^\dagger)^{-1}U_{\vec x-\hat\mu,\mu}^{-1}\right]\,.
\ee
The gauge cooling updates of the configuration are then given by
\begin{align}
U_{x,\hat\mu}&\mapsto \exp\left(-\sum_a \tilde\alpha \lambda_a
G_{a,x}\right) U_{x,\hat\mu}\,,\notag \\
U_{x-\hat\mu,\mu} &\mapsto U_{x-\hat\mu,\mu} \exp\left(\sum_a \tilde\alpha
\lambda_a G_{a,x}\right),
\label{gc}
\end{align}
where $\tilde\alpha=\epsilon \alpha$; $\alpha$  determines the strength of
the GC force, whereas $\epsilon$ is a discretization parameter
as in Eq.(\ref{dyn}). Note that even if $\tilde\alpha$ is not small,
Eq.(\ref{gc}) is still a gauge transformation; it just might not be
optimal for reducing $F$.

Gauge cooling first was shown to work in simple Polyakov loop model, given 
by a 1D lattice consisting of $N$ links with periodic boundary conditions. 
\cite{Seiler:2012wz} Analytically this model reduces to a trivial one-link 
integral, but it is a useful laboratory.
\be
-S = \beta_1 \tr U_1\ldots U_{N_t} +\beta_2 \tr U^{-1}_{N}\ldots
U^{-1}_1\,,
\ee
with $\beta_{1,2}$ complex. 
Simulating this model by using `uncooled' CL for the $N$ links, one 
discovers that already for $N=16$ one fails to reproduce the correct 
results. Adding sufficient gauge cooling, however, correct results are 
obtained. 

\section{GC for QCD and QCD inspired models}\label{sec-15}

\begin{center}
\begin{figure}[ht]
\includegraphics[width=0.8\columnwidth]{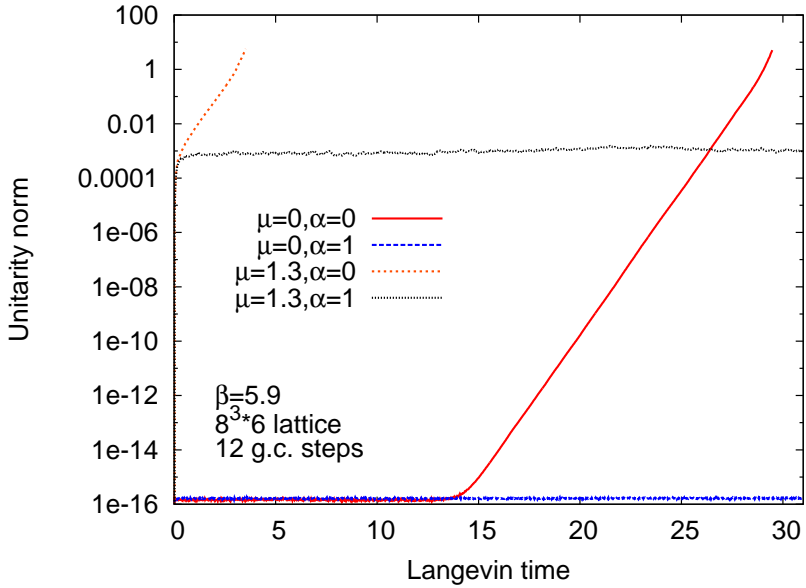} 
\caption{The unitarity norm in HDQCD as a function of Langevin time for 
various cooling parameters.}  
\label{unitnorm}
\end{figure}
\end{center}

A more interesting application of GC is the study of the so-called 
heavy-dense QCD (HDQCD) model, which is obtained by dropping all spatial 
links from the Wilson fermion action. The plot Fig.\ref{unitnorm} is from 
a study of this model \cite{Seiler:2012wz}.

\begin{center}
\begin{figure}[ht]
\includegraphics[width=0.8\columnwidth]{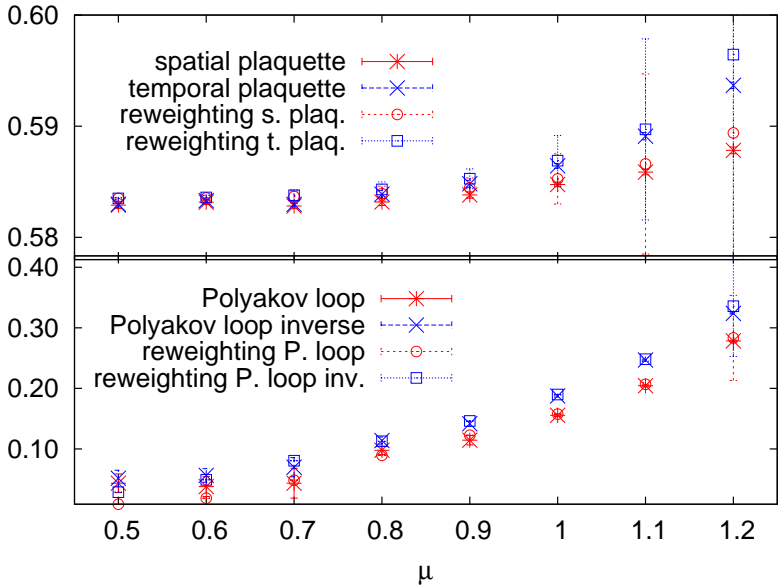}
\caption{HDQCD: comparison of CL results with reweighting.}
\label{hdqcd}
\end{figure}
\end{center}

This model had been studied in \cite{DePietri:2007ak} by reweighting, 
which is an exact procedure, but unfortunately limited to small lattices; 
here, however, those results serve as a standard to judge the correctness 
of CL simulations. In Fig.\ref{hdqcd} we show a comparison of the 
reweighted 
results with those of CL with GC on a $6^4$ lattice, taken from 
\cite{Seiler:2012wz}. It is seen clearly that for $\beta\ge 5.7$ there is 
excellent agreement, whereas for $\beta\le 5.6$ there are considerable 
deviations.

A natural explanation is that these deviations are due to poles in the 
drift at the zeroes of the determinant, leading to boundary terms 
or nonergodicity; just as in the toy models discussed before, increasing 
$\beta$ makes the process stay away from those dangerous places.

For full QCD there are also some results, some good, some not so good:

In an exploratory study for $\beta=5.9$ on a small ($4^4$) lattice Aarts 
et al \cite{Aarts:2014bwa} used the expansion in the spatial hopping 
parameter $\kappa_s$ for Wilson fermions, truncated at rather high order 
(up 50) to see convergence; the hopping parameters $\kappa_s=\kappa$ were 
chosen to be 0.12. Agreement of CL for the truncated expansion with CL 
for the full model supports correctness for this rather large value of 
$\beta$.

Sexty \cite{Sexty:2013ica} and later Aarts et al \cite{Aarts:2017vrv}
produced results for staggered quarks on lattices up to size $12^3\times 
4$; there are deviations at smaller $\beta$.

A very detailed study by Fodor et al \cite{Fodor:2015doa} compared CL with 
GC for QCD with staggered fermions to multi-parameter reweighting results 
on lattices up to $16^3\times 8$ and again found deviations for smaller 
$\beta$.

Kogut and Sinclair \cite{Sinclair:2016nbg,Sinclair:thiscontrib236} used 
staggered quarks on lattices up to $16^4$ at $\beta=5.6$ and found 
unphysical results: no clear `silver blaze' phenomenon and in particular 
wrong results for $\mu=0$, where the results can be checked by comparing 
with ordinary Monte Carlo simulations. By increasing $\beta$ to 5.7 that 
last discrepancy disappeared, in accordance with our general observation 
that larger $\beta$ values improve the situation. Their results show 
rather large values for the unitarity norms, which gives already reason 
for suspecting that the results are not correct (see next section).

Finally I should mention the very recent work by Bloch and Schenk 
\cite{Bloch:thiscontrib40} which is using a better matrix inversion method 
(`selected inversion') for the Dirac operator. This improves greatly 
the stability of the CL simulations, but brings out the deviations
for smaller $\beta$ even more clearly.

So this suggests that the incorrect results for QCD at smaller $\beta$ are 
not due to numerical instabilities, but rather to a more fundamental 
problem of CL, as discussed earlier in this talk.

Some more encouraging results for larger $\beta$ can be found, however, 
in \cite{Stamatescu:thiscontrib134}.

\section{Limitations of gauge cooling and how to deal with 
them}\label{sec-16}

We have seen that CL even with gauge cooling has some issues, which are 
quite likely due to the zeroes of the fermion determinant. Various cures 
for this problem have been tried: Nagata et al \cite{Nagata:2016vkn} 
suggest to use different unitarity norms, Bloch et al \cite{Bloch:2015coa} 
propose a variation of the GC procedure. In both cases problems at small 
$\beta$ remain. Bloch \cite{Bloch:2017ods} suggests reweighting of CL 
trajectories; while this works well in simple models, on larger lattices 
one has to expect overlap problems as always with reweighting.

\begin{center}
\begin{figure}[ht]
\includegraphics[width=0.8\columnwidth]{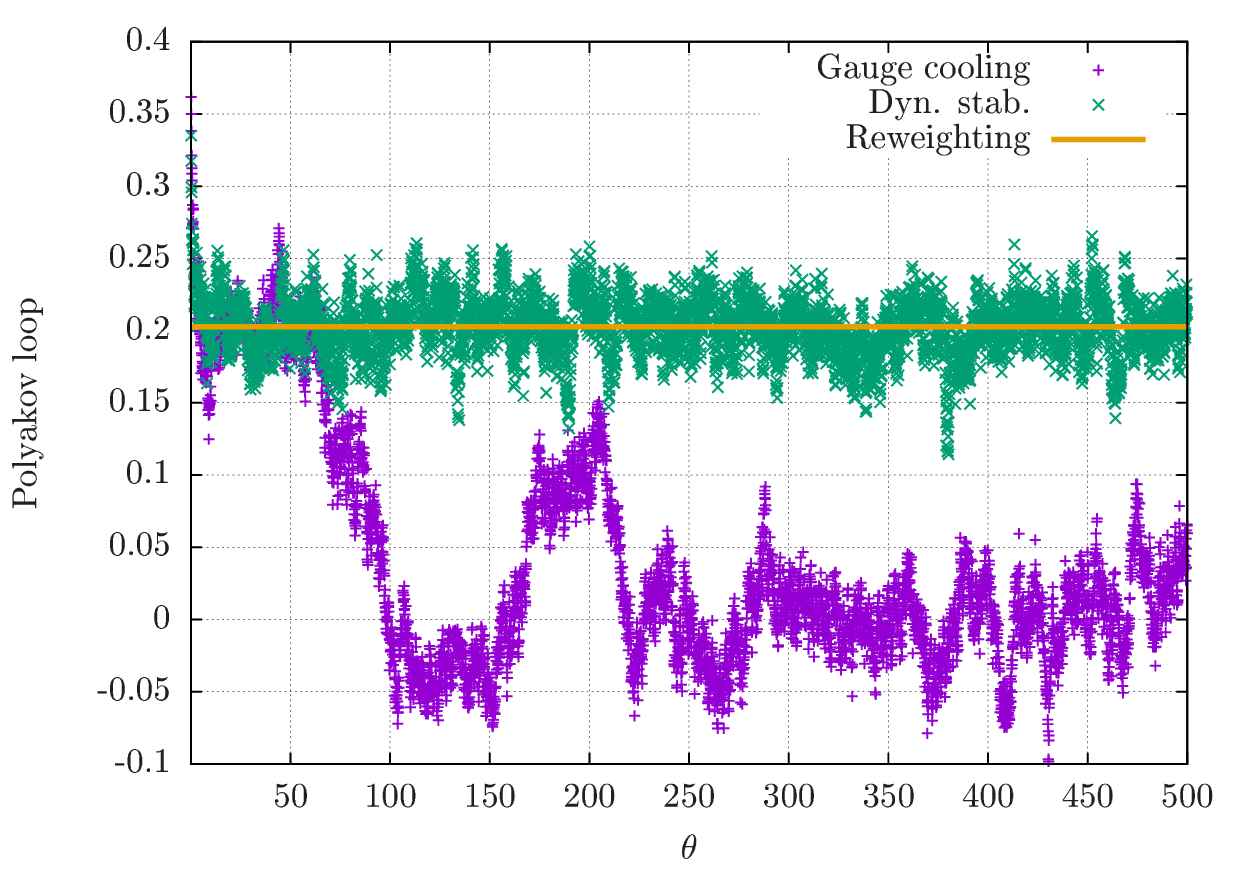}
\caption{HDQCD: time series of Polyakov loop vs Langevin time;
lattice $10^3\times 4, \kappa=0.04, \beta=5.8, \mu=0.7$.}
\label{ds}
\end{figure}
\end{center}

In my view the most promising, though not sufficiently understood cure is 
the so-called dynamical stabilization proposed by F.~Attanasio and 
B.~J\"ager \cite{Aarts:2016qhx,Attanasio:2016mhc,Jaeger:this contrib307}. 
The following plot Fig.\ref{ds} taken from\cite{Attanasio_thesis} and 
showing 
once more a simulation of HDQCD, gives some indication about what is going 
on: if we focus on the purple data, showing the evolution of a certain 
observable under CL with GC, we see that there is some kind of 
metastability up to $t\approx 70$, with the data fluctuating around the 
correct average value. Then the process wanders off fluctuating around a 
different (and incorrect) value.

A similar effect is also seen in the evolution of the unitarity norm: it 
first fluctuates around a small, but nonzero value, and then (again 
at $t\approx 70$) drifts off to a much larger value.

Dynamical stabilization is a somewhat ad hoc fix for this: it consists in 
the addition of a small extra drift $X$ keeping the unitarity norm small: 
\bea 
M^a_{\vec x}[U]&=ib^a_{\vec x}(b^a_{\vec x})^3\,,\quad 
b^a_{\vec x}=\sum_{\nu} \tr[\lambda^a 
U_{\vec x\nu} U_{\vec x\nu}^\dagger]\,, \notag\\ 
X^a_{\vec x\mu}&=i\epsilon\alpha_{DS} M^a_{\vec x}[U]\,. 
\eea 
$X$ is invariant under $SU(3)$ but not $SL(3,\C)$ gauge 
transformations. It is small as long as the unitarity norm remains in the 
right regime of fairly small values, and it is also small in the sense 
that it appears to go to zero with increasing $\beta$, indicating that it 
would disappear in the continuum limit, but it unfortunately invalidates 
the formal argument for correctness. For $\alpha=0$ it is absent, giving 
sometimes incorrect results as discussed, whereas for $\alpha=\infty$ it 
restricts the simulation to the unitary submanifold $\cM$ and is also 
incorrect. But by choosing $\alpha$ judiciously between those extremes, 
one can obtain good results.

In the picture Fig.\ref{ds} \cite{Attanasio_thesis} it is clearly seen 
that the extra drift makes the previously metastable region stable 
($\alpha$ was chosen appropriately). For details I refer to B.~J\"ager's 
contribution. \cite{Jaeger:this contrib307}.

It seems, however, necessary to gain a better understanding of the 
observed metastability. A possible interpretation is that, as in the 
$U(1)$ one-link model analyzed before, that the system crosses a 
bottleneck into the basin of attraction of another, `unphysical' 
attractive fixed point. Dynamical stabilization would then be just a 
device for preventing this crossing.

It is also conceivable that in order to stabilize the process, the 
unitary part of the gauge fluctuations needs to be restricted as well as 
the non-unitary one. Further research into these questions is necessary.

\section{Summary and outlook}\label{sec-17}

\begin{itemize}
\item
Insuffient decay both at $\infty$ and at poles may lead to boundary terms 
spoiling correctness of a CL simulation. Therefore `skirts' or `tails' 
have to be monitored carefully. 
\item
Gauge cooling eliminates some of the `skirts', but there still may be 
insufficient decay, so monitoring of them is still necessary.  
\item
Poles are harmless if the process stays away from them; monitoring this in 
QCD is costly, unfortunately. 
\item
The hopping expansion can sometimes avoid poles (see 
\cite{Aarts:2014bwa}), but it may still hit problems when $\kappa$ 
approaches the continuum value. 
\item
The best hope to cure the remaining problems seems to be the `dynamical 
stabilization', but a better understanding why this works seems necessary.
\item
In the context of the previous point, it seems highly desirable to gain a 
deeper understanding of the reason for the observed metastability in QCD, 
and more generally, understand the essentials of the fixed point and pole 
structure.

\end{itemize}

\bibliography{lattice2017}

\end{document}